\documentclass[12pt]{article}
\usepackage {epsf}
\usepackage{graphicx}
\usepackage{multirow}
\usepackage{slashed}
\usepackage{cite}
\usepackage {subfigure}
\usepackage {tabularx} 
\usepackage{rotate}
\usepackage[pdftex]{hyperref}

\newcommand{\bmat}{\left(\begin{array}}
\newcommand{\emat}{\end{array}\right)}
\def\NPB{Nucl. Phys. B}

\def\yzero{\smash{\hbox{$y\kern-4pt\raise1pt\hbox{${}^\circ$}$}}}

\def\a{\alpha}

\def\m{{\tilde m}}

\def\beq{\begin{equation}}
\def\eeq{\end{equation}}
\def\beqa{\begin{eqnarray}}
\def\eeqa{\end{eqnarray}}
\def\ba{\begin{array}}
\def\ea{\end{array}}

\def\-{\hphantom{-}}

\def\s2{\frac{1}{\sqrt2}}

\def\beq{\begin{equation}}
\def\eeq{\end{equation}}
\def\beqa{\begin{eqnarray}}
\def\eeqa{\end{eqnarray}}

\def\IF{\relax{\rm I\kern-.18em F}}
\def\II{\relax{\rm I\kern-.18em I}}
\def\IP{\relax{\rm I\kern-.18em P}}
\def\IC{\relax\hbox{\kern.25em$\inbar\kern-.3em{\rm C}$}}
\def\IR{\relax{\rm I\kern-.18em R}}

\def\cp{{\cal P}}

\def\Dsl{\,\raise.15ex\hbox{/}\mkern-13.5mu D} 
\def\IZ{Z\kern-.4em  Z}

 \def\cp#1{\relax\ifmmode {\IP\kern-2pt{}_{#1}}\else $\IP\kern-2pt{}_{#1}$\=fi}
\newcommand{\drawsquare}[2]{\hbox{%
\rule{#2pt}{#1pt}\hskip-#2pt
\rule{#1pt}{#2pt}\hskip-#1pt
\rule[#1pt]{#1pt}{#2pt}}\rule[#1pt]{#2pt}{#2pt}\hskip-#2pt
\rule{#2pt}{#1pt}}

\newcommand{\Ysymm}{\raisebox{-.5pt}{\drawsquare{6.5}{0.4}}\hskip-0.4pt%
        \raisebox{-.5pt}{\drawsquare{6.5}{0.4}}}
\newcommand{\Yasymm}{\raisebox{-3.5pt}{\drawsquare{6.5}{0.4}}\hskip-6.9pt%
        \raisebox{3pt}{\drawsquare{6.5}{0.4}}}

\catcode`\@=11
\newdimen\@rotdimen
\newbox\@rotbox

\def\@vspec#1{\special{ps:#1}}
\def\@rotstart#1{\@vspec{gsave currentpoint currentpoint translate
   #1 neg exch neg exch translate}}
\def\@rotfinish{\@vspec{currentpoint grestore moveto}}
\def\@rotr#1{\@rotdimen=\ht#1\advance\@rotdimen by\dp#1%
   \hbox to\@rotdimen{\hskip\ht#1\vbox to\wd#1{\@rotstart{90 rotate}%
   \box#1\vss}\hss}\@rotfinish}
\def\@rotl#1{\@rotdimen=\ht#1\advance\@rotdimen by\dp#1%
   \hbox to\@rotdimen{\vbox to\wd#1{\vskip\wd#1\@rotstart{270 rotate}%
   \box#1\vss}\hss}\@rotfinish}%
\def\@rotu#1{\@rotdimen=\ht#1\advance\@rotdimen by\dp#1%
   \hbox to\wd#1{\hskip\wd#1\vbox to\@rotdimen{\vskip\@rotdimen
   \@rotstart{-1 dup scale}\box#1\vss}\hss}\@rotfinish}%
\def\@rotf#1{\hbox to\wd#1{\hskip\wd#1\@rotstart{-1 1 scale}%
   \box#1\hss}\@rotfinish}%
\def\rotate{\@ifnextchar[{\@rotate}{\@rotate[l]}}
\def\@rotate[#1]#2{\setbox\@rotbox=\hbox{#2}\@nameuse{@rot#1}\@rotbox}

\catcode`\@=12

\renewcommand{\thefootnote}{\arabic{footnote}}
\begin{document}

\makeatletter \@addtoreset{equation}{section} 
\makeatother
\renewcommand{\theequation}{\thesection.\arabic{equation}}
\pagestyle{empty}
\pagestyle{empty}
\rightline{UOA/NPPS-07/04}
\vspace{0.01cm}

\begin{center}
\Large{{
Standard Models and Split Supersymmetry
from Intersecting Brane Orbifolds
}}
\\


{\large{ Christos Kokorelis 
}}
\\
\small{   Institute of Nuclear Physics, N.C.S.R. Demokritos, 
GR-15310, Athens, Greece\\
and\\
   Nuclear and Particle Physics Sector, Univ. of Athens,
GR-15771 Athens, Greece\\
and\\
Physics Division, National Technical University of Athens,
15780 Zografou Campus, Athens, Greece}\\[1mm]
\small{\bf Abstract} \\[1mm]
\end{center}

\begin{center}
\begin{minipage}[h]{16.0cm}
We construct four dimensional three generation 
non-supersymmetric $SU(3)_c \times SU(2)_L \times U(1)_Y$
intersecting D6-brane
models with $\nu_R$\rq{s}.
At three stacks we find exactly the MSSM chiral fermion matter spectrum.  
At 4-, 5-stacks we find models with the 
 massless fermion spectrum of the N=1 Standard Model and massive 
exotic non-chiral matter; these models flow also to only the SM.  
At 8-stacks 
we find MSSM-like models, with minimal massless exotics, made from two different N=1 sectors. Exotic triplet masses put a lower bound on the string scale of $2.79/2.89 \times 10^6$ GeV for a Higgs 124/126 GeV. It\rq{}s the first appearance of N=0 stringy quivers with the MSSM and matter in antisymmetric representations and perturbatively missing Yukawa couplings.  
The present models are based 
on orientifolds of ${\bf T^6/(Z_3 \times Z_3)}$ compactifications of 
IIA theory based on the torus lattice AAA; all complex moduli are fixed by the orbifold symmetry.   
We also present the spectrum rules + GS anomaly cancellation for the ABB lattice. 
Moreover, we point out the relevance of intersecting/and present D6-brane constructions on ideas related to existence of
split supersymmetry in nature.  
In this context we present non-susy models with only the SM-matter and also MSSM-matter dominated models, with massive gauginos and light higgsinos,  that achieve the correct 
supersymmetric GUT value for the Weinberg angle 
$sin^2 \theta = 
\frac{3}{8}$ at a string scale $5 \cdot 10^{13} \ GeV < M_{S}  <   1.4 \cdot 10^{17}$ GeV. It appears that if only the SM survives at low energy the unification scale is preserved at $5.03 \times 10^{13}$ GeV when n$_H$ =1, 3, 6. These models support the existence of split supersymmetry scenario in string theory.  
\end{minipage}
\end{center}

\newpage

\setcounter{page}{1} \pagestyle{plain}
\renewcommand{\thefootnote}{\arabic{footnote}}
\setcounter{footnote}{0}

\tableofcontents

\section{ Introduction}

Model building constructions (MBC\rq{}s) in the context of string theory have by far been 
explored both into the context of open string and heterotic string compactifications 
where a number of semirealistic have been
explored and analyzed \cite{pro1}. In the absence of a 
dynamical principle for selecting a particular string vacuum and 
simultaneously fix all moduli, the standard lore is to 
systematically analyze on phenomenological grounds 
the different string compactifications and trying to derive a miminal supersymmetric vaccum that may contain the Minimal Supersymmetric Standard Model (MSSM) if possible in the presence of a few exotics.  
Moreover, over the last few years, MBC\rq{}s coming from intersecting \footnote{We note that constructions with 
D6-branes intersecting at angles are T-dual to constructions with magnetic 
deformations \cite{ang, pra}, even though intersecting D-brane 
models has not yet been shown to be reproducible by the MD side.}
branes (IB's) \cite{lust1}-\cite{sca} 
have received a lot of attention as it become possible to construct 
 - for the
first time in string theory - 
non-supersymmetric (non-susy) four dimensional (4D) vacua with only 
the SM at low energy using intersecting 
 D6-branes
 \cite{louis2},\cite{kokos5}, \cite{kokos6}, \cite{suma} from 4D toroidal orientifolds of 
type IIA.  For other attempts to derive the SM from string theory see \cite{othera}. 
 In this regard, vacua based on non-SUSY 
Pati-Salam GUT constructions 
(with a stable proton), which break to the SM at low 
energy, giving masses to all exotics, have been also constructed and 
analyzed \cite{kokos1}.  
All the above models   
have vanishing RR tadpoles and uncancelled NS-NS 
tadpoles (coming from the closed string sector), the 
latter acting as an effective cosmological constant \cite{lust3}. 
On phenomenological grounds the string scale 
may be at the TeV; however as the D6-branes wrap the whole of internal space  
and there are no dimensions transverse to all branes, 
the presence of a TeV scale cannot be explained according
to the AADD mechanism \cite{savas}. \newline
Nevertheless, we note that intersecting brane worlds accommodate 
nicely the AADD \cite{savas} solution to the gauge hierarchy problem
by providing us with the
only known string realization in these backgrounds. 
Non-supersymmetric semirealistic GUTS in $Z_3$ 
orientifolds of interesting 
branes have been also analyzed for SU(5) \cite{lust3}. Flipped SU(5) GUTS were constructed \cite{nano} and the existence of appropriate Higgses necessary for the correct electroweak breaking to the SM at low energy and the doublet-triplet splitting mechanism 
has been shown \cite{axe}. The SU(5)/flipped SU(5) models are missing perturbatively the up/down quark couplings \footnote{For some attempts to derive the SM with quivers that are missing certain perturbative couplings but not 
based in a global string construction see \cite{toma}.}

Moreover, the construction of vacua which have only the MSSM at low energies,
has been also studied using either N=1 supersymmetric models or N=0
models that localize in part of their spectrum the MSSM.
In the latter case, the MSSM is localized as 
part of the non-supersymmetric 
open string spectrum where particles respect different N=1 supersymmetries \cite{iba, cre1} or where each particle of the MSSM preserves the same N=1 susy  and the rest of the spectrum a different N=1 susy\cite{kokosusy, floko}  
In the former case,
N=1 semirealistic supersymmetric vacua based on intersecting D6-branes 
has also been 
explored in four dimensional 
orientifolds of type IIA on ${\bf T^6/Z_2 \times Z_2}$ \cite{cve}, 
${\bf T^6/Z_4}$\cite{blu1}, 
${\bf T^6/Z_2 \times Z_4}$ \cite{gabi}, and ${\bf T^6/Z_6}$ \cite{ottho}, Z12-II \cite{bailin} and also N=1 GUT constructions have been analyzed \cite{cve}.
The main 
characteristics of these models is that not all 
complex structure moduli are fixed and  
part of their spectrum includes those 
of the N=1 SM (MSSM) in addition to extra massless chiral exotics \cite{cve}, 
\cite{ottho} or massless non-chiral exotics  \cite{blu1} [We also note that there 
are model building attempts from orientifolds of 
Gepner constructions where also the N=1 SM, with three pairs of $H_u$, $H_d$ MSSM Higgs 
multiplets, was found but in the presence of extra 
massless non-chiral exotics \cite{sche}.]
While supersymmetric models have no gauge hierarchy problem and are 
stable vacua as they do not have RR $\&$ NSNS tadpoles, we we will focus our attention to the MBC of  
non-supersymmetric models on 4D $Z3 \times Z3$ IIA orientifolds \cite{kokoneo} for several reasons.  
First of all, the satisfaction RR tadpoles and Green-Schwarz anomaly cancellation mechanism cancels all gauge anomalies .
Flavour changing neutral currents that could be a problem in low scale models $\sim 10^2 $ TeV \cite{abel} may be avoided as the scale of the models we study is at least $10^{13}$ GeV and higher. 
  There is no sign of any supersymmetry after the present 7 TeV run of LHC. At present there is no hint from squarks and gluinos below 1 TeV from their R-parity channels in the MSSM. Unification of of the three gauge couplings constants works
fine in the MSSM and does not work in the SM. In the context of intersecting branes \cite{gauge}, two of the gauge couplings unify, in a general N=1  supersymmetric model. Non-supersymmetric unification that we will study in this paper, is an open question.  
Complex moduli which can generate tadpoles are absent and the corresponding tadpoles vanish in the present models.
Thus the effect of the orbifold symmetries in the present models that fix complex structure moduli, is 
equivalent to the effects we achieve, by turning on arbitrary fluxes on three-cycles invariant under the discrete symmetry \cite{fluxes}.
Alternatively, one is using IIB to fix moduli \cite{bian}. 
 Only the dilaton has a only non-vanishing tadpole, which could cause an instability. However, this problem is unsolved and it is 
possible that the vacua re-adjust themselves so that the true vacuum is reached when all orders are taken into account in perturbation theory \cite{dienes}.

The purpose of this paper is to discuss the appearance of
N=0  
supersymmetric models that break to only the SM either without any 
exotics being present or with a minimal number of 
chiral \footnote{We do not present non-chiral exotics that are coming from vanishing intersections.} massive exotics [three (3) vector pairs]. 
These models are based
on four dimensional type IIA orientifolds on ${\bf T^6/{\bf Z_3 \times Z_3}}$
with D6-branes intersecting at angles \cite{kokoneo}.
In the N=0 intersecting D6-brane models 
presented in this work there are several interesting features :
\newline
a) Models which achieve the successful GUT result for the Weinberg angle, 
$sin^2 \theta = 3/8$ are presented.  
b) at the level of 3-stacks, we find models which break to only the SM at low 
energy. We also find non-susy models  - at 3- and 5-stacks - with
the chiral spectrum of the N=1 SM (with $\nu_R$'s) in the presence of three pairs of 
MSSM Higgsinos $H_u$, $H_d$ in addition to  
massive non-chiral exotics which again break to the SM at low energies.
A comment is in order.  
In this work when we will speak about the SM, we will keep in mind that in all
models there is no mass term for the up-quarks [The same effect persists
in the models of \cite{lust3, noti1}]. Instantons oud be recalled to generate the 
masses \cite{insta, ibaura}. \newline
Recently the split supersymmetry scenario (SS) was proposed \cite{split1}.
In this respect we propose  
intersecting D-brane models  that provide evidence for
a natural realization of the SS scenario in intersecting D-brane models as they
 satisfy most of the relevant criteria required by the SS existence.

The outline of the paper is as follows. 
In chapter 2, we will present the key features 
of the ${\bf T^6/{\bf Z_3 \times Z_3}}$ 
constructions, including the gauge 
group structure and spectrum 
rules. 
The details of the construction for the AAA, AAB, BBB lattice have been presented in a companion paper \cite{kokoneo}. 
In chapter 3, we discuss N=0 three generation (3G) non-supersymmetric SM\rq{}s
 with the fermion spectrum of the N=1 SM and extra non-chiral massive SU(3) triplet exotics. 
Higgsinos get massive and the models break to the SM at low energy. 
In chapter 4 we also discuss the deformation of these N=0 3G models to other N=0 
3G models which have only the SM the low energy without any exotics being present.
In chapter 5, we examine whether or not it is possible to 
construct N=0 models by using 4-stacks of D6's. 
In chapter 6 we present more possibilities for constructing N=0 3G models by using
five stacks of intersecting D6-branes. Here it is also possible to construct N=0 vacua 
with only the SM at low energy and extra non-chiral massive exotics. 
In chapter 7 we construct N=0 eight stack models with the N=1 SM fermionic spectrum made from N=1 SUSY preserving D6-branes.
In chapter 8, we present the spectrum rules for the $Z3 \times Z3$ on the ABB lattice. 
In chapter 9 we present arguments supporting the
 relevance of intersecting D-brane constructions 
to some new ideas related to the existence of split 
supersymmetry in nature and also discuss models with $sin^2 \theta =3/8$
at the string scale that satisfy most of the conditions required for the split susy
scenario.  Chapter 10, examines the unification of gauge couplings for the split susy models of Chapter 9 describing the SM unification from non-susy intersecting branes.
Chapter 11 contains our conclusions.

\section{\small Spectrum on $ T^6/Z_3 \times Z_3$ orientifolds, RR tadpoles 
$\&$ anomaly cancellation}

Our orientifold constructions originate from IIA theory compactified on the 
${\bf T^6/(Z_3 \times Z_3)}$ 
orbifold, where the latter symmetry is generated by the twist
generators (where $\a = e^{\frac{2\pi i}{3} }$)
$\theta : (z_1, z_2, z_3) \rightarrow (\a z_1, \a^{-1} z_2, z_3) $,  
$\omega :(z_1, z_2, z_3) \rightarrow ( z_1, \a z_2, \a^{-1}z_3)$,
where $\theta $, $\omega$ get associated to the twists 
$\upsilon =  \frac{1}{3}(1, -1, 0)$, 
$ u = \frac{1}{3}(0, 1, -1)$.  Here, $z^i = x^{i+3}+ i x^{i+5}$, 
$i=1,2,3$ are the complex coordinates on the  
$T^6$, which we consider as being factorizable for simplicity, e.g.
$T^6 = T^2 \otimes T^2 \otimes T^2$. 
In addition, to the orbifold action 
the IIA theory is modded out by the orientifold 
action $\Omega R$ that 
combines the worldsheet parity  $\Omega $ and the antiholomorphic operation 
$R : z^i \rightarrow {\bar z}^i$.
Because the orbifold action has to act crystallographically on the 
lattice the complex structure on all three $T^2$ tori is fixed to be
$U^I_A = 1/2 - i \sqrt{3}/2$.
The lattice vectors are defined as 
$e_1 = (1, 0)$, \ $e_2 = (-1/2, \sqrt{3}/2)$
and the 
$\Omega R $ action is along the horizontal directions across the six-torus. 
The model contains nine kinds of orientifold planes, that get associated to 
the orbit $\cal O$ consisting of the actions of $\Omega R$,  $\Omega R \theta$, $\Omega R \omega$,  
$\Omega R \theta^2$, $\Omega R \omega^2$, 
 $\Omega R \theta  \omega$,  $\Omega R \theta^2  \omega$, 
$\Omega R \theta  \omega^2$,  $\Omega R \theta^2  \omega^2$.
We will be interested on the open string spectrum and not discuss the closed
string spectrum that contains gravitational multiplets and orbifold 
moduli. In order to cancel the RR crosscap tadpoles introduced by the 
introduction of the orientifold planes we introduce N D6$_a$-branes wrapped
along three-cycles that are taken to be products of one-cycles along the 
three two-tori of the factorizable $T^6$. A D6-brane $a$ - 
associated with the equivalence class of wrappings $(n^I, m^I)$, $I=1,2,3$, - is mapped 
under 
the orbifold and orientifold action to its images
\beqa
a \leftrightarrow \left( \ba{c} n_a^1, m_a^1\\
 n_a^2 , \ m_a^2\\ n_a^3, \ m_a^3\ea \right), \ \theta a \rightarrow \left( \ba{c} 
-m_a^1 , \ (n-m)_a^1\\(m-n)_a^2 ,\ -n_a^2\\
n_a^3 ,\ m_a^3\ea \right), \ \ 
\Omega R a \rightarrow \left( \ba{c} (n-m)_a^1 ,\ -m_a^1\\
(n-m)_a^2 , \ -m_a^2\\ (n-m)_a^3 ,\ -m_a^3\ea \right) \ .
\eeqa

In $\Omega R$ orientifolds the twisted disk tadpoles vanish \cite{blumm}.
The $Z_3 \times Z_3$ orientifold models are subject to the cancellation of untwisted RR tadpole 
conditions \cite{kokoneo} given by
 \beq
\sum_a N_a  Z_a = 4, 
\label{tad}
\eeq    
where
\beq
Z_a = 2 m^1_a m^2_a m^3_a + 2 n^1_a n^2_a n^3_a - n_a^1 n_a^2 m_a^3 - 
n_a^1 m_a^2 n_a^3 - 
m_a^1 n_a^2 n_a^3 -
m_a^1 m_a^2 n_a^3 -m_a^1 n_a^2 m_a^3 - n_a^1 m_a^2 m_a^3 
\eeq
The gauge group U($N_a$) supported by $N_a$ coincident D6$_a$-branes comes 
from the $a({\tilde a})$ sector, the sector made from 
open strings stretched between the $a$-brane and its images under the orbifold 
action. In addition, we get three adjoint N=1 chiral multiplets. 
In the $a({\cal O}b)$ sector - strings stretched between the brane $a$ and the 
orbit images of brane $b$ - will localize $I_{ab}$ fermions in the 
bifundamental $(N_a, {\bar N}_b)$ where 
\beq
I_{ab}= 3(Z_a Y_b - Z_b Y_a),
\eeq 
and $(Z, Y) $ are the effective wrapping numbers with $Y_a$ given by
\beq
Y_a = m^1_a m^2_a m^3_a + n^1_a n^2_a n^3_a - n^1_a n^2_a m^3_a - 
n^1_a m^2_a n^3_a - m^1_a n^2_a n^3_a 
\eeq
The sign of $I_{ab}$ denotes the chirality of the associated fermion, where
we choose positive intersection numbers for left handed fermions. 
In the sector ${ab^{\prime}}$ - strings stretching between the brane $a$ and
the orbit images of brane $b$, there are $I_{ab^{\prime}}$ chiral fermions 
in the bifundamental $(N_a, N_b)$, with
\beq
I_{ab^{\prime}}= 3(Z_a  Z_b - Z_a Y_b - Z_b Y_a),
\eeq 
The theories also accommodate the following numbers of chiral fermions
in symmetric (S) and antisymmetric (A) representations of $U(N_a)$ from open
strings stretching between the brane $a$ and its orbit images $({\cal O}a)$, 
\beqa
(A_a) = 3(Z_a - 2 Y_a) ,\\
(A_a + S_a) = \frac{3}{2}(Z_a - 2 Y_a)(Z_a - 1)
\eeqa 
Finally, from open strings stretched between the brane $a$ and its orbifold 
images we get non-chiral massless fermions in the adjoint representation,
\beq
(Adj)_L : \prod_{i=1}^3 (L^I_{[a]})^2 \ ,
\label{adj1}
\eeq
where 
\beq
L^I_{[a]} = \sqrt{(m_a^I)^2 + (n_a^I)^2 -(m_a^I) (n_a^I) }
\label{nonma}
\eeq 
Adjoint massless matter, including fermions and gauginos that are massless at 
tree level 
are expected to receive string scale masses from loops once supersymmetry is broken 
\footnote{See the discussion in the appendix 
of \cite{louis2} and comments on section 8.}, leaving only the gauge bosons massless,
and we will not discuss it further.  
In the low energy theory, cubic gauge anomalies automatically cancel, 
due to the RR tadpole conditions (\ref{tad}). Mixed U(1)-gauge anomalies 
also cancel due to
the existence of a generalized Green-Schwarz (GS) mechanism 
(see \cite{kokoneo} for further 
details) that makes massive only one U(1) gauge field given by
\beq
\sum_a N_a (Z_a - 2 Y_a) F_a
\label{masi}
\eeq

The minimal choice of obtaining an extension of the 
Standard model (SM) is obtained using three 
stacks of D6-branes. The spectrum of open strings stretching 
between intersecting D6-branes 
is calculated by the use of rules (2.4 - 2.8). 
To establish notation we will denote the type of sypersymmetries 
preserved in the closed string sector by the choices of 
vectors \footnote{we follow the notation of the 
first reference of \cite{kokos1}.} $r_0 = \pm (1/2)(+-+-)$,  
$r_1 = \pm (1/2)(++--)$, $r_2 = \pm (1/2)(-++-)$, $r_3 = \pm (1/2)(----)$. 
Supersymmetry may be preserved by a system of branes if each stack of 
D6-branes is related to the $O6$-planes by a rotation in $SU(3)$, that
is the angles ${\tilde \theta}_i $ of the D6-branes with respect to 
the horizontal
direction in the i-th two-torus obeys the condition ${\tilde \theta}_1 + 
{\tilde\theta}_2 + {\tilde \theta}_3 = 0$. 
The supersymmetry of the 
models that is preserved by any pair of branes is determined by the 
choice of the orbifold and 
orientifold action. 
To examine whether N=0 or N=1
susy models are allowed we have to examine the brane wrappings $(n, m)$. Our current 
search finds no N=1 supersymmetric models but only N=0 ones, in chapter 8, made from different preserving supersymmetries.

\section{\small \em The N=0 MSSM}

Next, we obtain non-supersymmetric models which localize the fermion 
spectrum of the intersecting brane N=1 MSSM in addition to a couple of 
massive - non-chiral - exotics,  which subsequently
break to only the SM at low energy with the use of the GS mechanism described 
in the previous section.

\subsection{  N=0 SM's at low energy with the N=1 MSSM fermion spectrum at $M_s$}

The minimal choice of obtaining the SM gauge group and chiral spectrum 
is to start
from a three stack 
$U(3)_a \times U(2)_b \times U(1)_c$ D6-brane construction at the string scale.  The choice of wrapping numbers
\beq
(Z_a, Y_a) = \left( \ba{cc} 1,  & 0 \\\ea \right), \ (Z_b, Y_b) = 
\left( \ba{cc} 1, & 1 \\\ea \right), \ (Z_c, Y_c) = \left( \ba{cc}-1, & 1 \\\ea \right)
\label{wrap3}
\eeq 
satisfies the RR tadpoles and corresponds to the spectrum seen in 
table (\ref{taboldzero}). 

\begin{table}[htb] \footnotesize
\renewcommand{\arraystretch}{1.4}
\begin{center}
\begin{tabular}{|c||c||c||c||}
 \hline
 \multirow{1}{*}{{\bf Model structure}} & ${\bf Particles\hspace{1cm}}$ & ${\bf (SU(3) \times SU(2))_{(Q_a, Q_b, Q_c)}} \hspace{1cm}$ &  ${\bf U(1)^{Y}}$ \\\hline\hline
  \multirow{7}{*}{ $MSSM \ QUIVER$} & $\hspace{2cm} \{ Q_L \} \hspace{1cm}$   & $3({3}, { 2})_{(1,\ -1,\ 0)} \hspace{1cm}$ & $1/6$ \\
& $\hspace{2cm}\{ u_L^c  \}\hspace{1cm}$   & $\# \Yasymm \ = 3(3, 1)_{(2,\ 0,\ 0)}\hspace{1cm}$ & $-2/3$  \\
 & $\hspace{2cm}\{ d_L^c \}\hspace{1cm}$ & $3(3, 1)_{(-1,\ 0, \  -1)}\hspace{1cm}$ & $1/3$ \\
 & $\hspace{2cm}\{ L \ +  \ H_d \}\hspace{1cm}$ & $6(1, 2)_{(0,\  1,\  -1 )}\hspace{1cm}$ & $-1/2$ \\
 & $\hspace{2cm}\{ H_u \}\hspace{1cm}$ & $3(1, {\bar 2})_{(0,\  -1,\  -1 )}\hspace{1cm}$ & $1/2$ \\
 & $\hspace{2cm}\{ e_L^{+} \}\hspace{1cm}$ & $\# \bar{\Yasymm} \ = 3(1, 1)_{(0,\ -2, \ 0)}\hspace{1cm}$ & $1$    \\
 & $\hspace{2cm}\{ { N_R} \}\hspace{1cm}$ & $\# \Ysymm \ = 9(1, 1)_{({ 0}, \ { 0},\ { 2})}\hspace{1cm}$ & $0$ \\
 \hline\hline
  \multirow{2}{*}{Extra \ Matter}& $\hspace{2cm}\{ C_1 \}\hspace{1cm}$ & $3(3,\  1)_{(1, \ 0, \ -1 )}\hspace{1cm}$ & $1/3$ \\
 & $\hspace{3cm}\{ C_2 \}\hspace{2cm}$ & $3({\bar 3}, 1)_{(-1, \ 0, \ -1 )}\hspace{1cm}$ & $-1/3$ \\
\hline\hline
\end{tabular}
\end{center}
\caption{\small
A three generation non-supersymmetric model with the fermion content of 
N=1 MSSM on top of the table, in addition to $N_R$'s and three pairs
of $H_u$, $H_d$ Higgsinos. This model predicts nine $N_R$'s that could be written as a linear combination of three $N_R$\rq{}s. Note that this model
mimics models coming from gauge mediation scenarios.  
\label{taboldzero}}
\end{table}
We recognize in table (\ref{taboldzero}), the chiral spectrum
of the N=1 MSSM with three generations of right handed neutrinos ($N_R$)
and three pairs of massless `Higgsinos' \footnote{Instead of one Higgino $H_u$, $H_d$ 
pair in the standard 
global 
SUSY version of the MSSM.}. 
Also one U(1) gauge field 
becomes massive through its BF couplings, 
namely 
\beq
U(1)^{massive}=3F_a - 2F_b -3F_c
\label{mas1st}
\eeq
There hypercharge obeys the massleness condition
\beq
U(1)^{massive} \cdot U(1)^{massless} \ = 0
\label{condition} 
\eeq
thus surviving massless the GS mechanism (\ref{masi}).
The hypercharge reads:\newline
\beq
Y \ = \ U(1)^{Y} \ = \ -(1/3) F_a -(1/2)F_b
\label{choices} 
\eeq
The second U(1)\rq{}s that survive massless the (\ref{condition})  
\beq
U(1)^{add}= F_a -(2/3)F_b + (13/3)F_c \ .  
\eeq
$U(1)^{add}$ could be broken by a tachyonic singlet 
excitation charged under $F_c$, namely ${\tilde N}_R$,  
that plays the role of the `superpartner' of
$N_R$, thus leaving only the hypercharge $U(1)^{add}$ massless at low energies, below the
scale set by $ \langle {\tilde N}_R \rangle $.      
The following Yukawa couplings for the quarks, leptons and exotics $X_i$ 
are allowed:
\beqa
Y = \lambda_{d} \ Q_L d^c_L  {\tilde H}_d  {\tilde N}_R /M_s  +  
 \lambda_{u}^{ij} \ L^j  N_R^i  {\tilde H}_u \ + 
\lambda_{E} L E_R  
 {\tilde H}_d  {\tilde N}_R /M_s \ +  \ \nonumber\\
\lambda_{C} C_1 C_2  {\tilde N}_R \ + \lambda_{\mu} H_u  H_d {\tilde N}_R \ \ i=1,.,9; \ j=1,2
\label{fisu1}  
\eeqa 
The exotic triplets $C_i$ form a Dirac mass term 
which receives a mass of order
$M_s$ from the vev of ${\tilde N}_R$. 
The form of this coupling provide us with the bilinear mixing, that 
in a N=1 susy theory,  
would have played the role of a superpotential $\mu$-term.  
We remind that because the D6-branes wrap along all the $T^6$, the string 
scale is high and close to the Planck scale.
Thus the two Higgsinos $H_u$, $H_d$ receive a Dirac mass term from
the last term in (\ref{fisu1}) of order of the string scale, as the natural scale
of $\langle N_R \rangle  = M_s$. However, the    
value for the Higgsinos which can be at $M_s$ or lower is set by the
values of the Yukawa coupling coefficients $\lambda_{\mu}$. 
Large exponential suppression of a n-point interaction of Yukawa interactions in the form
\beq
\lambda_{\mu} \sim e^{-A}
\label{supr1}
\eeq 
is a natural aspect of IBW's due to their
dependence on the worldsheet area $A$, in string units,  located between their brane intersections 
\cite{abel, kokos1}. Hence a light higssino condensate of order of electroweak 
symmetry breaking $\upsilon = 246$ GeV can be obtained, assuming $M_s = 10^{16}$ GeV, 
with $A = 31$.

The quarks - apart for the u-quark which remain massless as the 
relevant coupling is excluded from charge conservation - and leptons  
receive non-zero masses from the Yukawa couplings in the 1st line of (\ref{fisu1}). 
Thus at low energy we have the SM - with the up quark remaining massless after electroweak 
symmetry breaking - and 
nine (9) generations of right handed neutrinos.
A comment is in order.
As the D6-branes involved wrap on generic 
angles the spectrum of table  (\ref{taboldzero}) is non-supersymmetric. 
Unfortunately, we were only able to find wrappings that render the models
non-supersymmetric.

\begin{table}[htb] \footnotesize
\renewcommand{\arraystretch}{1.4}
\begin{center}
\begin{tabular}{|c|c|}
\hline\hline
 $Brane$ & $\hspace{2cm}(n^1, m^1) \times (n^2, m^2) \times (n^3, m^3)$\hspace{2cm} \\
\hline\hline
$\{ a \}$   & $(1, \ 0) \times  (0,\ 1) \times (0,\ -1)$ \\
\hline 
$\{ b \}$   & $(0,\ 1)  \times (1,\ 0) \times (0, 1)$  \\
\hline 
$\{ c \}$   & $(0,\ -1)  \times (1,\ 1) \times (0, -1)$  \\
\hline
\hline
\end{tabular}
\end{center}
\caption{\small
Wrapping numbers responsible for the generation of the N=0 models of 
table (\ref{taboldzero}), that have the 
chiral spectrum of the intersecting brane N=1 MSSM.
\label{tabsu1} }
\end{table}
One can also check that the choice of effective wrappings
 \beq
(Z_a, Y_a) = \left( \ba{cc} 1, &  0  \\\ea \right), \ (Z_b, Y_b) = \left( \ba{cc} 1, & 1 
\\\ea \right), \ (Z_c, Y_c) = \left( \ba{cc}  -1, & -2 \\\ea \right)
\label{wrap33}
\eeq  
gives us also the N=0 chiral MSSM spectrum \footnote{Apart for some 
differences in the U(1) charges involved}  of table (\ref{taboldzero}) with the same 
hypercharge assignments.

\section{\bf Exactly the SM from three stacks}

In this section, we will construct non-supersymmetric models which have exactly the 
SM gauge group and chiral spectrum and no exotics present. These models will be  
constructed as a deformation of the models that appeared 
in table (\ref{taboldzero}). Also in these models there is no mass term
allowed from the up-quarks.

\subsection{ SM deformations of N=0 SM'S from three stacks}

$\bullet$ Let us make the choice of wrapping numbers
\beq
(Z_a, Y_a) = \left( \ba{cc} 1, & 0 \\\ea \right), \ (Z_b, Y_b) = \left(\ba{cc}1, 
&  1 \\\ea \right), \ (Z_c, Y_c) = \left( \ba{cc} -1, & -1 \\\ea \right)
\label{wrap341}
\eeq 

\begin{table}[htb] 
\caption{\small       
A three generation chiral (open string) spectrum accommodating the SM. 
The required Higgs may come from non-chiral bifundamentals in
the $bc$ sector \cite{louis2, kokos5, kokos6} that may trigger brane recombination. 
\label{tabold}}
\footnotesize
\renewcommand{\arraystretch}{2}
\begin{center}
\begin{tabular}{||r||c||c||}
\hline\hline
 ${\bf Matter\hspace{2cm}}$ & ${\bf (SU(3) \times SU(2))_{(Q_a, Q_b, Q_c)}} \hspace{2cm}$ & ${\bf U(1)^{Y}}$ \\
\hline\hline
$\hspace{2cm} \{ Q_L \} \hspace{2cm}$   & $3(3, {2})_{(1,\ -1,\ 0)} \hspace{2cm}$ & $1/6$  \\
\hline
$\hspace{2cm}\{ u_L^c \}\hspace{2cm}$   & ${\# \Yasymm}_a \ = 3({\bar 3}, 1)_{(2,\ 0,\ 0)}\hspace{2cm}$ & $-2/3$  \\
\hline
$\hspace{2cm}\{ d_L^c \}\hspace{2cm}$ & $3(3, 1)_{(-1,\ 0, \  1)}\hspace{2cm}$ & $1/3$ \\
\hline
$\hspace{2cm}\{ L \}\hspace{2cm}$ & $3(1, 2)_{(0,\  1,\  1 )}\hspace{2cm}$ & $-1/2$ \\
\hline
$\hspace{2cm}\{ e_L^{+} \}\hspace{2cm}$ & ${\# \bar{\Yasymm}}_b = 3(1, 1)_{(0, \ -2,\ 0)}\hspace{2cm}$ & $1$ \\
\hline
$\hspace{2cm}\{ N_R \}\hspace{2cm}$ & ${\# \bar{\Ysymm}}_c = 3(1, 1)_{(0, \ 0, \ -2 )}\hspace{2cm}$ & $0$ \\
\hline
\end{tabular}
\end{center}
\end{table}
This choice
satisfies the RR tadpoles and  
corresponds to the spectrum \footnote{The spectrum of this stringy quiver structure with matter in $\Ysymm$,  $\Yasymm$  representations has also appeared in the ${\IZ}_3$ orientifolds \cite{lust3}.}seen in 
table (\ref{tabold}). The intersection numbers are 
\beqa
I_{ab} = 3, & (A)_a = 3, &
I_{bc^{\star}} = 3, \nonumber\\ 
(A)_b = -3, &
I_{ac} = -3, &
(A+S)_c = -3 \ .
\eeqa
From (\ref{masi}) there is one anomalous U(1) which becomes massive
\beq
U(1)^{massive} = 3 F_a - 2 F_b + F_c
\eeq 
and two anomaly free U(1)'s  that correspond to the hypercharge and 
an extra U(1)
\beqa
U(1)^Y = -\frac{1}{3}F_a  -\frac{1}{2}F_b , \   \  \ 
U(1)^{ex}= -3 F_a + 2 F_b + 13 F_c 
\label{free}
\eeqa
We recognize in table (\ref{tabold}) exactly the chiral spectrum of 
the SM as at this point the spectrum for generic angles is 
non-supersymmetric. Exactly
the same quiver - but with opposite U(1) charges - non-supersymmetric 
chiral spectrum construction was found in \cite{lust3} from
intersecting D6-branes in $Z_3$ orientifolds. 
 In \cite{lust3} and in the present models the 
breaking of the extra U(1) surviving 
massless the Green-Schwarz mechanism proceeds via tachyonic excitations in the sector 
accommodating the right handed neutrino.
 Choices of
wrappings satisfying the constraints (\ref{wrap341}) can be 
seen in table (\ref{wrapthree}). Other choices of wrappings solving the RR tadpole
conditions may be seen
in table (\ref{newsu1}).
  Baryon (and lepton) number is not conserved but as 
the string scale in these models is naturally close to Planck scale we do 
expect a natural enhancement of gauge mediated proton decay modes and thus 
proton stability is guaranteed. 
The exchange of wrappings  
\beq
(Z_a, Y_a) \leftrightarrow (Z_b, Y_b) 
\label{exch1}
\eeq
is a symmetry of the theory as the spectrum and hypercharge 
of table (\ref{tabold}) do not 
change under the exchange (\ref{exch1}), which just reverses
the $U(1)_a$, $U(1)_b$ charges \footnote{obviously leaving invariant the hypercharge 
under field redefinition}.\newline
There is another symmetry under which the spectrum remains invariant.  
The spectrum remains invariant under the 
interchanges 
\beq
(n, m)_a \leftrightarrow (n, m)_b, \  \ (n, m)_a \leftrightarrow (n, m)_c, \
 \  (n, m)_b \leftrightarrow (n, m)_c \    
\label{chawra1}
\eeq  
applied in the wrappings of tables (\ref{wrapthree}), (\ref{newsu1}),  thus 
resulting in new N=0 
models. Some examples of this spectrum symmetry applied in the wrappings of 
table (\ref{newsu1}) may be seen in appendix A. 
\begin{table}[htb] \footnotesize
\renewcommand{\arraystretch}{2}
\begin{center}
\begin{tabular}{|c|c|}
\hline\hline
 $Brane$ & $\hspace{2cm}(n^1, m^1) \times (n^2, m^2) 
\times (n^3, m^3)$\hspace{2cm}\\
\hline\hline
$\{ a \}$   & $(1, \ 0) \times (0,\ 1)  \times (0,\ -1)$\\
\hline 
$\{ b \}$   & $(1,\ 1)  \times (1,\ 0) \times (-1, -1)$ \\
\hline
$\{ c \}$ & $(1,\ \ 1) \times (-1,\  0)  \times (-1,\ \ -1)$ \\
\hline
\hline
\end{tabular}
\end{center}
\caption{\small 
Wrapping numbers responsible for the generation of the three stack 
D6-brane non-supersymmetric Standard Models of table (\ref{tabold}). 
\label{wrapthree} }
\end{table}


\begin{figure}[t]
    \centering
    \includegraphics[width=0.8\textwidth]{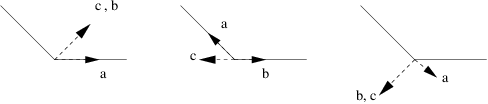}
    \caption{Brane positions in the SM's of table (\ref{tabold}) for the wrapping choices 
of table (\ref{wrapthree}). }
    \label{fig:awesome_image} 
\end{figure}

The Higgs available for electroweak symmetry breaking (ESB) may come 
from bifundamental 
scalars that may be understood as part of the massive N=2 hypermultiplet spectrum 
containing also the massive scalars 
stretched 
\begin{table}[htb] \footnotesize
\renewcommand{\arraystretch}{2}
\begin{center}
\begin{tabular}{|c|c|}
\hline\hline\hline\hline
 $Brane$ & $\hspace{2cm}(n^1, m^1) \times (n^2, m^2) 
\times (n^3, m^3)$\hspace{2cm} \\
\hline\hline
$\{ a \}$   & $(1, \ 0) \times (0,\ 1)  \times (0,\ -1)$\\
\hline 
$\{ b \}$   & $(1,\ 1)  \times (1,\ 0) \times (-1, -1)$ \\
\hline
$\{ c \}$ & $(1,\ \ 1) \times (1,\  1)  \times (1,\ \ 0)$ \\
\hline
\hline
\end{tabular}
\end{center}
\caption{\small 
Wrapping numbers responsible for the generation of the three stack 
non-supersymmetric 
D6-brane Standard Models of table (\ref{tabold}).
\label{newsu1} }
\end{table}
between the U(2) brane and the brane image of the U(1); 
as the b,c branes are parallel in at least
one complex plane along the different orbits.  
The Higgs scalars 
become tachyonic [see also \cite{louis2, kokos5, kokos6}] 
by varying the distance between 
the parallel branes.  
The available electroweak Higgs have the quantum 
numbers
\beq
 h_1 = (1,2)_{(0,\ 1, \ -1)}, \
h_2 = (1, {\bar 2})_{0,\  -1, \ 1)} \ ,
\eeq
where the allowed Yukawa couplings are given by 
\beq
Y = \lambda_d Q_L d_L^c h_1 \  + \ \lambda_{\nu} L N_R h_2  \ +  \ \lambda_e L E_R h_1 
\eeq
with no mass term for u-quarks.

$\bullet$$\bullet$ {\bf Other examples of SM\ wrappings} A three stack N=0 model with the 
chiral spectrum of only the SM can be also derived from the wrapping numbers 
(\ref{wrap341}), by 
deforming around the $Y_c$ wrapping number.
Thus the choice of wrappings
 \beq
(Z_a, Y_a) = \left( \ba{cc}1 , & 0 \\
\ea \right), \ (Z_b, Y_b) = \left( \ba{cc} 1, & 1 \\ \ea \right), \ (Z_c, Y_c) 
= \left( \ba{cc} -1, & 0 \\\ea\right)
\label{wrap300}
\eeq    
provide us with the spectrum of table (\ref{tabold}) but with reversed 
$U(1)_c$ 
charge. 
In this case the U(1) gauge field which 
becomes massive through its nonzero
coupling to the RR fields is given by $3F_a - 2F_b - F_c$. Also the  
hypercharge and the extra U(1) are given respectively by  
\beq
U(1)^Y = -\frac{1}{3}F_a - \frac{1}{2}F_b, \ \ U(1)^{add} = 
-\frac{3}{13}F_a +\frac{2}{13}F_b - F_c
\eeq
A set of wrappings associated with the effective wrappings (\ref{wrap300}) 
is given in table (\ref{newset1}). A different set of wrappings solving
the RR tadpoles  
may be seen in table (\ref{newset100}). Notice that for the wrapping numbers of the a, b branes seen in tables (\ref{newset1}), (\ref{newset100}), the ab-intersection that localizes $Q_L$ doublets, preserves 
the N=1 supersymmetry $r_2$. Thus we have a similar effect (as the one appearing in \cite{kokos5} where the  SM $\&$ $\nu_R$\rq{}s appears and in addition the tadpole solutions allow the spartners of the right handed neutrinos to exist in a N=1 supersymmetric intersection in an overall non-supersymmetric model) where the spectrum of the SM with right handed neutrinos of table (\ref{tabold}) is overall non-supersymmetric but the Quark doublet is N=1 supersymmetric. 
\begin{table}[htb] \footnotesize
\renewcommand{\arraystretch}{2}
\begin{center}
\begin{tabular}{|c|c|}
\hline\hline
 $Brane$ & $\hspace{2cm}(n^1, m^1) \times (n^2, m^2) \times (n^3, m^3)\hspace{2cm}$ \\
\hline\hline
$\{ a \}$   & $(1, \ 0) \times (0,\ 1)  \times (0,\ -1)$ \\
\hline 
$\{ b \}$   & $(1,\ 1)  \times (1,\ 0) \times (-1, -1)$  \\
\hline
$\{ c \}$   & $(0,\ \ -1) \times (0,\  1)  \times (-1,\ \ 0)$ \\
\hline
\hline
\end{tabular}
\end{center}
\caption{\small
Wrapping numbers in the three stack non-supersymmetric D6-brane Standard Models 
of the wrapping choices (\ref{wrap300}) of table (\ref{tabold} with reversed U(1)$_c$ charge. 
The ab intersection preserves the $r_2$ susy. 
\label{newset1}}
\end{table}
One can also check that the wrapping solutions of table (\ref{newset1})  
have the S3 permutational symmetry (\ref{chawra1}) that allows the individual pairs of wrappings 
$(n^i, m^i$) of the $T^2_i$  
tori to permute with the wrappings of the other $T^2_i$, $i \neq j$, tori.     
 We also note that in these models there is no mass term for the up-quarks as well.  
\begin{table}[htb] \footnotesize
\renewcommand{\arraystretch}{2}
\begin{center}
\begin{tabular}{|c|c|}
\hline\hline
 $Brane$ & $\hspace{2cm}(n^1, m^1) \times (n^2, m^2) \times (n^3, m^3)$\hspace{2cm}  \\
\hline\hline
$\{ a \}$   & $(1, \ 0) \times (0,\ 1)  \times (0,\ -1)$ \\
\hline 
$\{ b \}$   & $(1,\ 1)  \times (1,\ 0) \times (-1, -1)$ \\
\hline
$\{ c \}$ & $(0,\ \ 1) \times (0,\  1)  \times (1,\ \ 0)$ \\
\hline
\hline
\end{tabular}
\end{center}
\caption{\small
Wrapping numbers responsible for the 
non-supersymmetric three stack D6-brane Standard Models of the wrapping 
choices (\ref{wrap300}) of table (\ref{tabold} with reversed U(1)$_c$ charge. The ab intersection preserves the $r_2$ susy.
\label{newset100} }
\end{table}

\section{ \bf Four stacks of 
D6-branes and massive exotics}
 
In this section, we will exhibit the appearance of three generation
 non-supersymmetric models  
by using four stacks of D6-branes.
We will not give a very detail description of these models as the issue 
of whether the SM gauge group survives massless to low energies.   
We are
considering a system of four stacks of D6-branes, namely we 
start with a 
gauge group $U(3)_a \times U(2)_b \times U(1)_c \times U(1)_d$ at the string 
scale $M_s$.

\subsection{ The N=0 Standard Models}

We choose the effective wrappings 
\beqa
(Z_a, Y_a) = \left( \ba{cc} 1 , & 1 \\\ea \right) , \  
(Z_b, Y_b) = \left( \ba{cc}  1 , & 0 \\\ea \right), \nonumber\\
(Z_c, Y_c) = \left(  \ba{cc} 1, & 1 \\\ea  \right), \ 
 (Z_d, Y_d) = \left(\ba{cc} -2 , &  -2  \\\ea  \right) \ .
\label{wrap} 
\eeqa
which  satisfies the RR tadpole conditions (\ref{tad}).
For the
choice of wrappings seen in table (\ref{wrasu10}) the 
models are  
non-supersymmetric.
The chiral spectrum of the set of effective wrappings (\ref{wrap}) which is associated
with the  
hypercharge assignment $U(1)^Y = (1/3)F_a  + (1/2)F_b $
may be seen in table (\ref{newta}).
\begin{table}[htb] \footnotesize
\renewcommand{\arraystretch}{1.5}
\begin{center}
\begin{tabular}{|c|c|c|}
\hline\hline
 $Brane$ & $\hspace{2cm}(n^1, m^1)(n^2, m^2)(n^3, m^3)$\hspace{2cm}  \\
\hline\hline
$\{ a \}$   & $(1, \ 1)(1,\ 0)(-1,\ -1)$  \\
\hline
$\{ b \}$   & $(1,\ 0)(0,\ 1)(0, -1)$  \\
\hline
$\{ c \}$ & $(1,\ \ 1)(1,\ 0) (-1,\ \ -1)$  \\
\hline
$\{ d \}$ & $(1,\ 1)(1,\  0)( 2, \ 0 )$ \\
\hline
\hline
\end{tabular}
\end{center}
\caption{\small
Wrapping numbers in the four stack D6-brane Standard-like Models that are generated 
by the choice of 
effective wrappings (\ref{wrap}).
\label{wrasu10} }. 
\end{table}
\begin{table}[htb]
 \footnotesize
\renewcommand{\arraystretch}{1.4}
\begin{center}
\begin{tabular}{||c||c||c||l||}
\hline
 \multirow{1}{*}{{\bf Model structure}} & ${\bf Particles}$ & ${\bf U(1)^{Y}}$ &${\bf (SU(3) \times SU(2))_{(Q_a, \ Q_b, \ Q_c,\  Q_d)}}$  \\\hline\hline
 \multirow{7}{*}{$MSSM \ QUIVER$} & $\{ Q_L \} $ &  $1/6$  & $3({\bar 3}, {2})_{(-1, \ 1,\  0,\  0)} $  \\
& $\{ u_L^c \}$ &  $-2/3$  & $3({\bar 3},1)_{(-2,\ 0,\ 0, \ 0)}$  \\
& $\{ d_L^c \}$ & $1/3$ &$3(1, 1)_{(1, \ 0,\ 0,\ 1 )}$ \\
& $\{ L \}$ & $-1/2$ & $3(1, 2)_{(0, \ -1,\ 0, \ 1 )}$ \\
& $\{ H_u \}$ & $1/2$ & $3(1,{2})_{(0, \ 1,\ -1, 0 )}$\\
& $\{ H_d \}$ & $-1/2$ & $3(1,{2})_{(0, \ -1,\ 0,\ 1 )}$ \\
& $\{ e_L^{+} \}$ & $1$ & $3(1, 1)_{(0, \ 2,\ 0,\  0)}$ \\
& $\{ S_1 \equiv \nu^c_L \}$ & $0 $  & $3(1, 1)_{(0, \ 0,\ 0,\ -2 )}$\\
\hline
\hline
\multirow{2}{*}{ $ Singlets$}& $\{ S_2 \}$ & $0$ & $6(1, 1)_{(0,\  0,\ 0, \ -2)}$ \\
& $\{ S_3 \}$ &  $ 0  $   & $3(1, 1)_{(0, \ 0,\ 1, \ 1 )}$\\
\hline\hline
\multirow{2}{*}{ $ Exotics$}& $\{ X_1 \}$ & $-1/3$ & $3({\bar 3}, 1)_{(-1,\  0,\ -1, \ 0)}$ \hspace{2cm} \\
&$\{ X_2 \}$ & $1/3$ & $3(3, 1)_{(1,\  0,\ 0,\ 1 )}$\\
\hline
\end{tabular}
\end{center}
\caption{\small  
The three generation N=0 SM-like models from four stacks of intersecting branes 
with its chiral spectrum with three pairs of Higgsinos and right handed 
neutrinos. 
{\em Either one of the gauge multiplets $S_I$ could be 
identified as the one
associated with the right handed 
neutrino}. The exotics triplets $X_I$ receive a Dirac mass  
only with the choice of hypercharge associated to $Y^1$. The scalar massive 
superpartners of the 
singlets $S_2$, 
$S_3$ -  after they become tachyonic - may be used 
to break the extra U(1)'s. These N=0 SM-like models have the symmetry under the 
exchange $d_L^c   \leftrightarrow X_2$ , $ L  \leftrightarrow H_d $.
\label{newta} }
\end{table}

The Yukawa couplings  are 
\beqa
Y_{(table \ \ref{newta})} \ = \ 
\lambda_d Q_L d_L^c H_d^H S_2^H /M_s \ 
+ \ \lambda_e L e_L^{+}H_d^H S_2^H /M_s + \ 
\lambda_{\nu} L \nu_L^c H_u^H S_3^H /M_s \ + \ &\nonumber\\ 
 \lambda_{\mu}^{(4)} H_u H_d S_3^H S_2^H /M_s  \  + \
 \lambda_{(12)} {X}_1 {X}_2  S_3^H S_2^H /M_s  \ ,
\label{firstw}
\eeqa
where by $H_d^H$,  $H_u^H$, $S_3^H$, $S_2^H$ we denote the massive 'superpartners' of 
the matter  $H_d$,  $H_u$, $S_3$, $S_2$ respectively.
The  $H_u H_d$,  ${X}_1 {X}_2$ exotic pairs form Dirac mass terms respectively, 
that receive a non-zero mass
from the combined effect of the 
vevs of the scalar superpartners of
$S_2$, $S_3$  which become tachyonic.
All SM fermions but the one associated to $u_L^c$ 
transform in bifundamentals. Moreover the Yukawa couplings   
give masses to all quarks and leptons but the u-quark, for which the
relevant term is excluded from charge conservation.

We however note that under the brane recombination (BR) ${\tilde c} = c + d$, the 
four 
stack models of table (\ref{newta}) flow to the non-supersymmetric three stack 
models of table (\ref{BRfir1}). 

\begin{table}[htb] \footnotesize
\renewcommand{\arraystretch}{2}
\begin{center}
\begin{tabular}{||r||c||c||}
\hline\hline
 ${\bf Matter\hspace{2cm}}$ & ${\bf (SU(3) \times SU(2))_{(Q_a, Q_b, Q_c)}} \hspace{2cm}$ & ${\bf U(1)^{Y}}$ \\
\hline\hline
$\hspace{2cm} \{ Q_L \} \hspace{2cm}$   & $3({\bar 3}, {2})_{(-1,\ 1,\ 0)} \hspace{2cm}$ & $1/6$  \\
\hline
$\hspace{2cm}\{ u_L^c \}\hspace{2cm}$   & $3({\bar 3}, 1)_{(-2,\ 0,\ 0)}\hspace{2cm}$ & $-2/3$  \\
\hline
$\hspace{2cm}\{ d_L^c \}\hspace{2cm}$ & $3(3, 1)_{(1,\ 0, \  -1)}\hspace{2cm}$ & $1/3$ \\
\hline
$\hspace{2cm}\{ L \}\hspace{2cm}$ & $3(1, 2)_{(0,\  -1,\  1 )}\hspace{2cm}$ & $-1/2$ \\
\hline
$\hspace{2cm}\{ e_L^{+} \}\hspace{2cm}$ & $3(1, 1)_{(0, \ 2,\ 0)}\hspace{2cm}$ & $1$ \\
\hline
$\hspace{2cm}\{ N_R \}\hspace{2cm}$ & $3(1, 1)_{(0, \ 0, \ -2 )}\hspace{2cm}$ & $0$ \\
\hline
\hline
$\hspace{2cm}\{ S_0 \}\hspace{2cm}$ & $6(1, 1)_{(0, \ 0, \ -2 )}\hspace{2cm}$ & $0$ \\
\hline
\end{tabular}
\end{center}
\caption{\small
A three generation non-supersymmetric chiral (open string) spectrum accommodating the SM that result
from brane recombination on the SMs of table (\ref{newta}). The extra singlet becomes massive
by its coupling to the tachyonic scalar superpartner. The latter may be also be used to break the 
extra U(1), beyond hypercharge, surviving massless the Green-Schwarz mechanism.   
\label{BRfir1}}
\end{table}
We also note that the wrappings coming from the interchanges (\ref{chawra1}) 
are still a symmetry of the spectrum.

\section{\bf Only the (N=0) MSSM from five stacks of D6-branes}

In this section, we will investigate the possibility to construct N=0
models by using a higher numbers of stacks, namely 
five stack vacua. In these models  
the SM will survive massless below the string scale to low energies. 
After the Green-Schwarz anomaly 
cancellation the N=0 models will localize the massless fermion spectrum of the 
N=1 SM, which in turn will be reduced with the help of Higgs tachyons to that of only 
the SM at low energies.
The five stack configuration involves the initial localization of 
chiral models with a
$U(3)_a \times U(2)_b \times U(1)_c \times U(1)_d \times U(1)_e $  gauge group 
at the string scale.

These N=0 models are constructed from the effective wrapping numbers
\beqa
(Z_a, Y_a)  = \left( \ba{cc}
 1 \ ,&  1\\  \ea \right) ,  \  \     (Z_b, Y_b) = \left( \ba{cc}
    1\ ,& 0 \\  \ea   \right),  \  \  
(Z_c, Y_c)  = \left(\ba{cc} 1 , &  1 \\\ea  \right),\nonumber\\  
 (Z_d, Y_d) = \left( \ba{cc} -1\ , &  -1 \\  \ea    \right),   \  \  
(Z_e, Y_e) = \left( \ba{cc}  -1 , &  -1  \ea \right) \ .  
\label{kowra1}
\eeqa
The above choice of wrapping numbers satisfies the RR tadpole cancellation condition 
(\ref{tad}). The corresponding three generation chiral spectrum can be seen in 
table (\ref{newta0}). 
 \begin{table}[htb]
 \footnotesize
\renewcommand{\arraystretch}{1.4}
\begin{center}
\begin{tabular}{||c||c||c||l|}
\hline
 \multirow{1}{*}{{\bf Model structure}} & ${\bf Particles}$ & ${\bf U(1)^{Y}}$ &${\bf (SU(3) \times SU(2))_{(Q_a, \ Q_b, \ Q_c,\  Q_d, \ Q_e)}}$  \\\hline\hline
 \multirow{7}{*}{$MSSM \ QUIVER$} & $\{ Q_L \} $ &  $1/6$  & $3(3, {\bar 2})_{(-1, \ 1,\  0,\  0, \ 0)} $  \\
& $\{ u_L^c \}$ &  $-2/3$  & $3({\bar 3},1)_{(-2,\ 0,\ 0, \ 0, \ 0)}$\hspace{2cm}  \\
& $\{ e_L^{+} \}$ & $ 1 $ & $3(1, 1)_{(0,\ 2,\ 0, \ 0,\  0)}$\hspace{2cm} \\
& $\{ d_L^c \}$ & $1/3$ & $3(1, 1)_{(1, \ 0,\ 0,\ 1, \ 0 )}$ \hspace{2cm} \\
& $\{ H_u \}$ & $1/2$ & $3(1, 2)_{(0, \ 1,\ -1, \ 0,\ 0 )}$ \hspace{2cm} \\
& $\{ L \}$ & $-1/2$ & $3(1,{2})_{(0, \ -1,\ 0, \ 1, \ 0 )}$\hspace{2cm} \\
& $\{ H_d \}$ & $-1/2$ & $3(1,{2})_{(0, \ -1,\ 0,\ 0, \ 1 )}$ \hspace{2cm} \\
& $\{ S_3 \equiv \nu_L^c \}$ & $0$ & $3(1, 1)_{(0, \ 0,\ 1,\ 1, \ 0)}$\hspace{2cm} \\
\hline\hline
\multirow{4}{*}{ $Extra \ singlets$}& $\{ S_1 \}$ & $0$ & $3(3, 1)_{(0,\  0,\ 0,\ -2, \ 0  )}$\hspace{2cm} \\
& $\{ S_2 \}$ & $0$ & $3(3, 1)_{(0,\  0,\ 0,\ 0, \ -2 )}$\hspace{2cm} \\
& $\{ S_4  \}$ & $0 $  & $3(1, 1)_{(0, \ 0,\ 1,\ 0,\ 1 )}$\hspace{2cm} \\
& $\{  S_5 \}$ & $ 0 $   & $6(1, 1)_{(0, \ 0,\ 0,\ -1, \ -1 )}$ \hspace{2cm} \\
\hline
\hline
\multirow{2}{*}{ $ Exotics$}& $\{ X_1 \}$ & $-1/3$ & $3({\bar 3}, 1)_{(-1,\  0,\ -1, \ 0, \ 0)}$ \hspace{2cm} \\
& $\{ X_2 \}$ & $1/3$ & $3(3, 1)_{(1,\  0,\ 0,\ 0, \ 1 )}$\hspace{2cm} \\
\hline
\end{tabular}
\end{center}
\caption{\small 
On the top of the table the N=0 models with the three generation MSSM 
chiral spectrum with three pairs of Higgsinos and right handed neutrinos. 
{\em Either one of the gauge multiplets $S_I$ could be 
identified as the one
associated with the right handed 
neutrino}. The exotics triplets $X_I$ form a Dirac mass term, leaving
only the SM at low energy.  
\label{newta0} }
\end{table}

The analysis of U(1) anomalies in the models shows that there is a 
massive U(1) 
given by the combination $U(1)^{(1)}= -3 F_a + 2 F_b - F_c + F_d + F_e$ and 
also another 
four U(1)'s - including the hypercharge - 
which survive massless the Green-Schwarz mechanism, namely the 
following 
\beqa
U(1)^{(2)} = \frac{1}{3} F_a + \frac{1}{2} F_a \ ,\  \ \ U(1)^{(3)}= 3 F_a - 2 F_b - 13 F_c 
&\nonumber\\
U(1)^{(4)}= 3 F_a - 2 F_b + F_c + 7 F_d + 7 F_e \ , \ \  \ U(1)^{(5)}= F_d - F_e \ .
\label{fifthsta1}
\eeqa
The extra U(1)'s may be broken by the vevs of vevs of the superpartners of the 
 $S_1$, $S_2$, $S_4$, $S_5$, namely the  $S_1^H$, $S_2^H$, $S_4^H$, $S_5^H$,
 Thus for example $S_4^H$ may be used to break $U(1)^{(3)}$, 
$S_5^H$ may be used to break $U(1)^{(5)}$, while $S_1^H$, $S_2^H$ could be 
used to break $U(1)^{(4)}$.  Thus at low energies only the SM gauge group 
survives.

We construct N=0 models with the spectrum 
of table (\ref{newta0}). A choice of wrappings can be seen in 
table (\ref{susy1}).
Further examples of wrappings which describe equivalent models can be seen in
 appendix C, in tables 
(\ref{susy2}), (\ref{susy3}) and (\ref{susy4}). These models of 
appendix C are constructed by the application of the interchange 
of wrappings - the latter being a 
symmetry of the
spectrum - in (\ref{chawra1}) to the wrappings of table (\ref{susy1}).    
The models of table 
(\ref{susy1}) are non-susy \footnote{We follow the notation
of the first reference of \cite{kokos1}.}.

\begin{table}
[htb] \footnotesize
\renewcommand{\arraystretch}{1.4}
\begin{center}
\begin{tabular}{|c|c|}
\hline\hline
 $Brane$ & $\hspace{2cm}(n^1, m^1)(n^2, m^2)(n^3, m^3)$\hspace{2cm} \\
\hline\hline
$\{  a \}$   & $(1, \ 1)(1,\ 0)(-1,\ -1)$  \\
\hline
$\{ b \}$   & $(1,\ 0)(0,\ 1)(0, -1)$   \\
\hline
$\{ c \}$ & $(1,\ \ 1)(1,\ 0) (-1,\ \ -1)$   \\
\hline
$\{ d \}$ & $(1,\ 1)(1,\  0)( 1, \ 1 )$ \\
\hline
$\{ e \}$ & $(1,\ 1)(1,\  1)( 1, \ 0 )$   \\
\hline
\hline
\end{tabular}
\end{center}
\caption{\small 
Wrapping numbers responsible for
N=0 supersymmetry in the five stack 4D three generation intersecting D6-brane models.
\label{susy1} }
\end{table}
We note that in all models there is no mass term for the up-quarks which 
is excluded from charge
conservation.

$\bullet$ {\bf Model A}

Yukawa couplings for the quarks and leptons and exotic 
triplets $X_i$ of the models appearing in table (\ref{newta0}) 
are given by 
\beqa
Y^{(table \ \ref{newta0})}\ =  \ 
 \lambda_{d} \frac{Q_L d_L^c H_d^H  S_4^H }{M_s} \ +  \
\lambda_{e} \frac{L e_L^{+} H_d^H S_5^H }{M_s} 
\ + \ \lambda_{\nu} \frac{L \nu_L^c H_u^H S_1^H }{M_s}
\ +  \ \nonumber\\
 \lambda_x X_1 X_2 S_4^H \ + \  H_u H_d \frac{(   \lambda_{\mu}^{(2)} S_3^H S_5^H +  \lambda_{\mu}^{(1)} 
S_4^H S_2^H)}{M_s}  \nonumber\\
\label{supw1}
\eeqa
The mass term for the exotic triplets couples to the vev of the 
superpartner of $S_4$.
As the presence of these triplets can mediate scalar mediated
proton decay modes such as the
\beq
(  ({\bar u}_{L}^c)_{\alpha} (d_{L})_{\beta} ) 
( ({\bar d}_L^c)_{\gamma} \nu_L ) \epsilon_{\alpha \beta \gamma}\ ,
\eeq
it is necessary that they receive a mass which it is 
at least $10^{16}$ GeV or higher,
such as proton decay is enhanced beyond the 
observable present limit $\Gamma^{-1}_{expected}$
($p \rightarrow e^{+}_L \pi_o) \geq 10^{33}$ yrs [ See also \cite{igorwi} 
for the 
calculation of proton decay rate for a general N=1 SU(5) model in the context of 
$Z_2 \times Z_2$ orientifolds and also \cite{axe} for complementary considerations on stringy 
proton decay and doublet-triplet splitting for SU(5) and flipped SU(5) GUTS ].
Hence,
it is guaranteed that the scalar mediated proton decay modes are suppressed.  
The chiral fermions $H_u$, $H_d$, receive a Dirac mass from the last term  
in eqn. (\ref{supw1}).

$\bullet$ {\bf Model B}

An alternative class of N=0 supersymmetric models, where also
 all exotics are massive, can be derived from the models appearing 
in table (\ref{newta0}) by the exchanges
\beq
L \leftrightarrow H_d, \ \ \ d_L^c \leftrightarrow X_2,
\label{ret1}
\eeq
which can be obviously be chosen due to the degeneracy of their hypercharge. 
The spectrum of the new models can be seen in table (\ref{newtala0}).  
The Yukawa couplings for quarks, leptons and exotics colour triplets $X_i$ are
 \begin{table}
[htb] \footnotesize
\renewcommand{\arraystretch}{1}
\begin{center}
\begin{tabular}{|c|c|c|}
\hline\hline
\  \ \  \ ${\bf Matter\  \ for \  \ Y^1}$ & \ $Y^1$ 
\ & $(SU(3) \times SU(2))_{( Q_a, \ Q_b,\ Q_c,\ Q_d, \ Q_e )}\hspace{2cm}$  \\
\hline\hline
$\{ Q_L \}$  & $1/6$ & $3(3, {\bar 2})_{(-1, \ 1, \ 0, \ 0, \ 0)}$\hspace{2cm}  \\
\hline
$\{ u_L^c \}$ &  $-2/3$  & $3({\bar 3},1)_{(-2,\ 0,\ 0, \ 0, \ 0)}$\hspace{2cm}  \\
\hline
$\{ e_L^{+} \}$ & $ 1 $ & $3(1, 1)_{(0,\ 2,\ 0, \ 0,\  0)}$\hspace{2cm} \\
\hline
$\{ d_L^c \}$ & $1/3$ & $3(1, 1)_{(1, \ 0,\ 0,\ 0, \ 1 )}$ \hspace{2cm} \\
\hline
$\{ H_u \}$ & $1/2$ & $3(1, 2)_{(0, \ 1,\ -1, \ 0,\ 0 )}$ \hspace{2cm} \\
\hline
$\{ H_d \}$ & $-1/2$ & $3(1,{2})_{(0, \ -1,\ 0, \ 1, \ 0 )}$\hspace{2cm} \\
\hline
$\{ L \}$ & $-1/2$ & $3(1,{2})_{(0, \ -1,\ 0,\ 0, \ 1 )}$ \hspace{2cm} \\
\hline-
$\{ S_3 \equiv \nu_L^c \}$ & $0$ & $3(1, 1)_{(0, \ 0,\ 1,\ 1, \ 0)}$\hspace{2cm} \\
\hline\hline
$\{ S_1 \}$ & $0$ & $3(3, 1)_{(0,\  0,\ 0,\ -2, \ 0  )}$\hspace{2cm} \\
\hline
$\{ S_2 \}$ & $0$ & $3(3, 1)_{(0,\  0,\ 0,\ 0, \ -2 )}$\hspace{2cm} \\
\hline
$\{ S_4  \}$ & $0 $  & $3(1, 1)_{(0, \ 0,\ 1,\ 0,\ 1 )}$\hspace{2cm} \\
\hline
$\{  S_5 \}$ & $ 0 $   & $6(1, 1)_{(0, \ 0,\ 0,\ -1, \ -1 )}$ \hspace{2cm} \\
\hline
\hline
$\{ X_1 \}$ & $-1/3$ & $3({\bar 3}, 1)_{(-1,\  0,\ -1, \ 0, \ 0)}$ \hspace{2cm} \\
\hline
$\{ X_2 \}$ & $1/3$ & $3(3, 1)_{(1,\  0,\ 0,\ 1, \ 0 )}$\hspace{2cm} \\
\hline
\end{tabular}
\end{center}
\caption{\small 
The three generation N=0 SM from from five stacks of intersecting branes 
with its chiral spectrum and three pairs of Higgsinos.
On the top of the table the chiral structure of N=1 SM. The middle 
part exhibits the gauge singlets while the bottom 
part includes the triplet exotics. These models can come from the models of 
table (\ref{newta0}) by the exchange (\ref{ret1}). At low energy only the SM survives.
\label{newtala0} }
\end{table}
\beqa
Y_{(table \ \ref{newtala0})}\ =  \ 
 \lambda_{d} \frac{Q_L d_L^c H_d^H  S_5^H}{M_s} \ +  \ 
\lambda_{e} \frac{L e_L^{+} H_d^H S_5^H}{M_s } 
\ + \ \lambda_{\nu} \frac{L \nu_L^c H_u^H S_5^H}{M_s}  \ + \ &\nonumber\\
  X_1 X_2 \frac{(\lambda_x^{(1)} S_3^H S_1^H + \lambda_x^{(2)} S_4^H S_5^H)}{M_s} \ .
\label{supw10}
\eeqa
We observe that there is a universality in the dependence of the mass terms for the 
down quark, the electron and the neutrino mass
on the vev of the $S_5$ previously massive superpartner. The latter  
Higgs tachyonic field 
can generate natural mass scales of the electroweak order in the 
following 
sence.
Take for example the mass for the d-quark. Its mass is given by $m_d = \lambda_d u_d$, where 
$\langle H_d^H \rangle = \upsilon_d$. Thus the required hierarchy for the mass of 
the d-quark, 
$m_d^{exp} = 0.05$ GeV, may be generated from the exponential suppression 
generated by the
Yukawa coupling factor $ \lambda_d$ of the relevant four point function.
The fermions $ H_u$, $H_d$ receive a non-zero mass from the Yukawa interaction terms   
\beq 
H_u H_d (\lambda_{\mu}^{(1)}  \frac{S_3^H S_1^H}{M_s} + \lambda_{\mu}^{(2)} \frac{S_4^H S_5^H}{M_s})
\eeq

$\bullet$ {\bf Model C}

Another interesting class of N=0 supersymmetric models, where also
 all exotics are massive, can be derived from the models of
 table (\ref{newta0}) by the exchanges
\beq
L \leftrightarrow H_d
\label{reta0}
\eeq
These models are further analyzed in appendix C.

$\bullet$ {\bf Model D}

A further N=0 3G 4D model, with the chiral spectrum of the intersecting brane N=1 SM at the string scale, 
 is 
obtained by the exchange
\beq
d_L^C \leftrightarrow X_2
\label{exapen1}
\eeq
on the particle spectrum of table (\ref{newta0}). 
These models are examined in appendix D.

\begin{itemize}

\item Brane recombination

\end{itemize}

The string theory recombination process (BR) 
should be better described by 
string field theory. For some examples with BR involving classical methods 
at the level of gauge theory, see \cite{nagaoka}. 
In the present models, BR works as follows : 
a) Under the BR ${\tilde c} = c + d + e$, the 5-stack models of table 
(\ref{newtala0}), flow to the three stack models of table (\ref{flotab1}).     

\begin{table}[htb] \footnotesize
\renewcommand{\arraystretch}{1.5}
\begin{center}
\begin{tabular}{|r|c|c|}
\hline\hline
 ${\bf Matter\hspace{2cm}}$ & $(SU(3) \times SU(2))_{(Q_a, Q_b, Q_{\tilde c})} \hspace{2cm}$ & $U(1)^{Y}$ \\
\hline\hline
$\hspace{2cm} \{ Q_L \} \hspace{2cm}$   & $3({\bar 3}, {2})_{(-1,\ 1,\ 0)} \hspace{2cm}$ & $1/6$  \\
\hline
$\hspace{2cm}\{ u_L^c \}\hspace{2cm}$   & $3({\bar 3}, 1)_{(-2,\ 0,\ 0)}\hspace{2cm}$ & $-2/3$  \\
\hline
$\hspace{2cm}\{ d_L^c \}\hspace{2cm}$ & $3(3, 1)_{(1,\ 0, \  1)}\hspace{2cm}$ & $1/3$ \\
\hline
$\hspace{2cm}\{ L \}\hspace{2cm}$ & $3(1, 2)_{(0,\  -1,\  1 )}\hspace{2cm}$ & $-1/2$ \\
\hline
$\hspace{2cm}\{ e_L^{+} \}\hspace{2cm}$ & $3(1, 1)_{(0, \ 2,\ 0)}\hspace{2cm}$ & $1$ \\
\hline
$\hspace{2cm}\{ N_R \}\hspace{2cm}$ & $3(1, 1)_{(0, \ 0, \ -2 )}\hspace{2cm}$ & $0$ \\
\hline
\hline
$\hspace{2cm}\{ S_0 \}\hspace{2cm}$ & $3(1, 1)_{(0, \ 0, \ -2 )}\hspace{2cm}$ & $0$ \\
\hline
\end{tabular}
\end{center}
\caption{\small
A three generation non-supersymmetric chiral (open string) spectrum accommodating the SM that 
comes
from brane recombination on the 5-stack SM with massive exotics of table (\ref{newta}). 
\label{flotab1}}
\end{table}

b) Under the BR ${\tilde d} = d + e$, the 5-stack models of table 
(\ref{newtala0}), flow to the 4-stack models of table (\ref{flotab2}).

\begin{table}[htb] \footnotesize
\renewcommand{\arraystretch}{1.5}
\begin{center}
\begin{tabular}{|r|c|c|}
\hline\hline
 ${\bf Matter\hspace{2cm}}$ & $(SU(3) \times SU(2))_{(Q_a, Q_b, Q_c, Q_{\tilde d})} 
\hspace{2cm}$ & $U(1)^{Y}$ \\
\hline\hline
$\hspace{2cm} \{ Q_L \} \hspace{2cm}$   & $3({\bar 3}, {2})_{(-1,\ 1,\ 0)} \hspace{2cm}$ & $1/6$  \\
\hline
$\hspace{2cm}\{ u_L^c \}\hspace{2cm}$   & $3({\bar 3}, 1)_{(-2,\ 0,\ 0)}\hspace{2cm}$ & $-2/3$  \\
\hline
$\hspace{2cm}\{ X_1 + d_L^c \}\hspace{2cm}$ & $6(3, 1)_{(-1,\ 0, \  0, \  -1)}\hspace{2cm}$ & $1/3$ \\
\hline
$\hspace{2cm}\{ X_2 \}\hspace{2cm}$ & $3({\bar 3}, 1)_{(-1,\ 0, \  -1, \  0)}\hspace{2cm}$ & $-1/3$ \\
\hline
$\hspace{2cm}\{ H_d + L \}\hspace{2cm}$ & $6(1, 2)_{(0,\  -1,\  0, \ 1 )}\hspace{2cm}$ & $-1/2$ \\
\hline
$\hspace{2cm}\{ H_u \}\hspace{2cm}$ & $3(1,2)_{(0,\  1,\  -1, \ 0 )}\hspace{2cm}$ & $1/2$ \\
\hline
$\hspace{2cm}\{ e_L^{+} \}\hspace{2cm}$ & $3(1, 1)_{(0, \ 2,\ 0)}\hspace{2cm}$ & $1$ \\
\hline
$\hspace{2cm}\{ S_1 \}\hspace{2cm}$ & $6(1, 1)_{(0, \ 0, \ 1, \ 1 )}\hspace{2cm}$ & $0$ \\
\hline
\hline
$\hspace{2cm}\{ S_2 \}\hspace{2cm}$ & $6(1, 1)_{(0, \ 0, \ 0,  -2 )}\hspace{2cm}$ & $0$ \\
\hline
\end{tabular}
\end{center}
\caption{\small
A three generation non-supersymmetric chiral (open string) spectrum accommodating the SM 
that comes
from brane recombination on the 5-stack SM with massive exotics of 
table (\ref{newta}).   
\label{flotab2}}
\end{table}

\section{\bf Three generation N=0 MSSM-like models made from N=1 supersymmetric D6-branes}

\subsection{An example of a N=0 MSSM-like model}

\begin{figure}
    \includegraphics[scale=0.8]{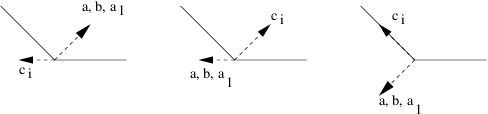}
\caption{Brane positions in the non-supersymmetric MSSM-like models of table (\ref{susymodels}) for the wrapping choices 
of the set solution $(l_4, s_3)$; a-brane $\equiv U(3)_a$, b-brane $\equiv U(2)_b$, a$_1$-brane $\equiv U(1)_c$. The N=0 models are made from two different N=1, N$^{\prime}$=1 supersymmetry preserving sectors. The wrappings can be seen in table (\ref{taneto}).}
\label{figd}
\end{figure}

Let us consider the eight stack N=0 model that satisfies the RR tadpole conditions 
\beq
3 Z_a \ + \ 2 Z_b \ + \ Z_c 
\ + \sum_{ i=1}^5 Z_{c_i}  \ = \ 4
\eeq
with its effective wrappings given by  
\beqa
(Z, Y)_a &=& (-1, \ -1), \ (Z, Y)_b = (-1, \ -1), \ (Z, Y)_{a_1} = (-1,\ -1),   \nonumber\\
 (Z, Y)_{c_i} &=& (2, \ 1), \  i = 1,\cdots,5 .
\label{mstad}
\eeqa
The initial gauge group is based on the structure  
$U(3)_a \times U(2)_b   \times U(1)_{a_1} 
  \times U(1)_{c_1}   \times U(1)_{c_2}   \times U(1)_{c_3}
  \times U(1)_{c_4}  \times U(1)_{c_5}$.
The full chiral spectrum of these N=0 models can be seen in 
table (\ref{susymodels}). Regarding the 
wrappings $(n, m)$ associated to $(Z, Y)$, we find different solutions for the $(Z, Y) \leftrightarrow (n, m)$ wrappings that are associated with the preservation of  N=1 supersymmetry on a \lq\lq{}hidden\rq\rq{} single $U(1)_{c_i}$ D6-brane and a different N=1 supersymmetry on the \lq\lq{}observable\rq\rq{} sector MSSM branes. 
They are listed in table (\ref{neo1}). 
We classify the effective wrappings (Z, Y) by the set $(l_j, s_i)$ for convenience.
\begin{table}
[htb] \footnotesize
\renewcommand{\arraystretch}{1}
\begin{center}
\begin{tabular}{|c|c|c|c|}
\hline\hline
Solution & $(Z, \ \ Y)$ & $\hspace{1cm}(n^1,\ \  m^1)(n^2,\ \  m^2)(n^3,\ \m^3)$\hspace{1cm} 
& N=1 SUSY \ Preserved\\
\hline\hline
$l_1$ & $ (-1,\ \ -1)$   & $(1,\ \  1)(1,\ \ 0)(1,\ \  1)$  & $r_1, r_2$ \\
\hline
$l_2$ & $ (-1,\ \ -1)$   & $(1,\ \  0)(1,\ \ 1)(1,\ \  1)$  & $r_1, r_0$ \\
\hline
$l_3$ & $ (-1,\  \ -1)$   & $(1,\ \  1)(1,\ \ 1)(1,\ \ 0)$  & $r_2, r_0$ \\
\hline
$l_4$ & $ (-1,\ \ -1)$   & $(1,\ \ 1)(-1,\ \ 0)(-1,\ \  -1)$  & $r_2$ \\
\hline
$l_5$ & $ (-1,\ \ -1)$   & $(-1,\ \ 0)(1,\ \ 1)(-1,\ \  -1)$  & $r_2$ \\
\hline
$l_6$ & $ (-1,\ \ -1)$   & $(-1,\ \ -1)(-1,\ \ 0)(1,\ \ 1)$  & $r_1$ \\
\hline
$l_7$ & $ (-1,\ \ -1)$   & $(-1,\ \ -1)(-1,\ \ -1)(1,\ \  0)$  & $r_0, r_2$ \\
\hline
$l_8$ & $ (-1,\ \ -1)$   & $(-1,\ \ -1)(1,\ \ 0)(-1,\ \  -1)$  & $r_1, r_2$  \\
\hline
$l_9$ & $ (-1,\ \ -1)$   & $(1,\ \ 0)(-1,\ \ 0)(1,\ \ 1)$  & $-$  \\
\hline
\hline
$s_1$ & $ (2,\ \  1) $   & $(1, \ \ 1)(0,\ \ 1)(-1,\ \ 0)$ & $r_3$  \\
\hline
 $s_2$   & $ (2,\ \ 1) $   & $(0, \ \  1)(-1,\ \ 0)(1,\ \ 1)$ & $r_3$  \\
\hline
 $s_3$   & $ (2,\ \ 1) $   & $(-1,\ \ 0)(1,\ \ 1)(0,\ \ 1)$ & $r_3$  \\
\hline
$s_4$ & $ (2,\ \ 1) $   & $(1, \  \ 1)(0,\ \ -1)(1,\ \ 0)$ &  $r_3$  \\
\hline
 $s_5$   & $ (2,\ \ 1) $   & $(0, \ \  -1)(1,\ \ 1)(1,\ \ 0)$ & $r_3$  \\
\hline
 $s_6$   & $ (2,\ \ 1) $   & $(1, \ \ 0)(0,\ \ -1)(1,\ \  1)$ & $r_3$  \\
\hline
$s_7$ & $ (2,\ \ 1) $   & $(0, \ \ 1)(-1,\ \ -1)(1,\ \ 0)$ & $r_3$  \\
\hline
 $s_8$   & $ (2,\ \ 1) $   & $(-1, \ \ -1)(0,\ \ 1)(1,\ \  0)$ &  $r_3$  \\
\hline
 $s_9$   & $ (2,\ \ 1) $   & $(1,\ \ 0)(0,\ \ -1)(1,\ \ 1)$ &  $r_3$ \\
\hline\hline\hline
\end{tabular}
\end{center}
\caption{\small 
Multiple solutions for the wrapping numbers of the eight stack 4D N=0 three generation 
intersecting D6-brane models.  
\label{neo1}}
\end{table}
For example, by using the wrapping numbers 
of the pairs $(l_i, s_j )$ for the (observable, hidden) sector branes respectively,  as seen in table (\ref{taneto}),  to be e.g. all equal to the $(l_4, s_3)$ wrappings respectively, we can construct a non-supersymmetric model which accommodates an \lq\lq{}observable\rq\rq{} MSSM sector respecting a single N=1 supersymmetry and is made of $l_4$ associated D6-branes and a \lq\lq{}hidden\rq\rq{}  sector
which preserves a (different) $N^{\prime}=1$ supersymmetry and is made from the $s_3$ D6-branes. These wrappings can be seen in figure (\ref{figd}). The exchange symmetries (\ref{chawra1}) are also valid in these models.  
 A similar construction based on 4D toroidal orientifolds of type IIA \cite{lust1} has appeared in \cite{floko}. 
\begin{table}
[htb] \footnotesize
\renewcommand{\arraystretch}{1}
\begin{center}
\begin{tabular}{|c|c|c|c|}
\hline\hline
Solution/branes  & $(Z, \ \ Y)$ & $ (n^1,\ \  m^1)(n^2,\ \  m^2)(n^3,\ \m^3)$ 
& N=1 SUSY \\
\hline\hline
\lq\lq{}Observable\rq\rq{} : $l_4 = a, b, a_1$ & $ (Z, \ Y) = (-1,\ \ -1)$   & $(1,\ \  1)(-1,\ \ 0)(-1,\ \  -1)$  & $r_2$ \\
\hline
\lq\lq{}Hidden\rq\rq{} : $s_3 = c_1, c_2, c_3, c_4, c_5$ & $ (Z, \ Y) = (2,\ \ 1)$   & $(-1,\ \  0)(1,\ \ 1)(0,\ \  1)$  & $r_3$ \\
\hline
\hline\hline
\end{tabular}
\end{center}
\caption{\small 
Wrapping numbers of the 8-stack 4D N=0 three generation 
intersecting D6-brane models. The \lq\lq{}observable\rq\rq{} MSSM stacks of $l_4$ branes preserve a different N=1 supersymmetry than the universal N$^{\prime}$ =1 supersymmetry preserved by the extra $s_3$ \lq\lq{}hidden\rq\rq{} branes. 
\label{taneto}}
\end{table} 


\begin{itemize}

\item U(1) anomalies

\end{itemize}
The analysis of U(1) anomalies shows that there is one U(1) which becomes 
massive though its couplings to RR fields, namely
\beq
U(1)^{mas} = 3F_a + 2F_b + F_{a_1} \ , 
\eeq   
while the hypercharge which remains massless is given 
by $(1/3) F_a -(1/2)F_b $. There is also a third $U(1)^c =(3/2)F_a + F_b 
-(13/2)F_{a_1}$ 
which can be broken by the vev of the tachyon singlet superpartner of $S_0$, and also five more 
U(1)'s which are linear combinations of all five U(1)'s, 
$U(1)^{c1}, \cdots, U(1)^{c5}$ and can be broken e.g. by the vev's of one of 
the singlet tachyonic superpartners of $S_1, \cdots, S_5$. As there are more tachyonic 
singlets available in the 
models e.g. $S_1, \cdots, S_9$, there are 
different choices of singlets that could be used to break the extra, beyond the 
hypercharge, surviving massless the Green-Schwarz mechanism U(1)'s.

\begin{itemize}

\item Chiral Spectrum

\end{itemize}

Most of the matter becomes massive by appropriate
Yukawa couplings - denoted in table (\ref{susymodels}) by using the \rq\rq{}+\rq\rq{} 
sign.  
There are only \footnote{, apart from the $up$-quarks in the $\bar{ \Yasymm}$ of SU(3) which are massless,}
  two pairs of chiral fields    
where the matter in each pair has opposite hypercharges with respect to the 
the surviving gauge group $SU(3)_c \times SU(2)_L \times U(1)_Y$ - that we 
were not able to
find a  mass term;  
denoted in table (\ref{susymodels}) by using the \rq\rq{}-\rq\rq{} sign.   
However, even with the latter drawback it is worthwhile to examine some of 
its revealing phenomenology.
\begin{table}
\renewcommand{\arraystretch}{0.65}
\begin{center}
\begin{tabular}{||r||c||c||c||}
\hline\hline
 ${\bf Matter}\hspace{1cm}$ & $Massive$       &  ${(SU(3) \times SU(2))}_{(Q_a, Q_b, Q_{a1},Q_{c1}, 
Q_{c2}, Q_{c3},Q_{c4}, Q_{c5} )} $ 
& $U(1)^{Y}$ \\
\hline
$ \{ Q_L \}\hspace{1cm} $ & $+$ & $3({3^{*}}, { 2^{*}})_{(-1,\ -1,\ 0,\ 0,\ 0,\ 0,\ 0,\ 0)} 
$ & $1/6$  \\
\hline
$\{ u_L^c  \}\hspace{1cm}$ & $-$ & $3({\bar 3}, 1)_{(-2,\ 0,\ 0,\ 0,\ 0,\ 0,\ 0,\ 0)}$ & $-2/3$  \\
\hline
$\{ d_L^c \}\hspace{1cm}$ &  $+$  & $3(3, 1)_{(1,\ 0, \ 0, \ -1,\ 0,\ 0, \ 0,\ 0)}$ & $1/3$ \\
\hline
$\{ H_u \}\hspace{1cm}$ & $+$ & $3(1, {2^{*}})_{(0,\  -1,\  -1,\ 0,\ 0,\ 0,\ 0,\ 0)}$ & $1/2$ \\
\hline
$\{ H_d^2 \equiv H_d \}\hspace{1cm}$ & $+$ & $3(1, {2})_{(0,\ 1,\ 0,\ 1,\ 0,\ 0,\ 0,\ 0)}$ & $-1/2$ \\
\hline
$\{ e_L^{+} \}\hspace{1cm}$ & $+$ & $3(1, 1)_{(0, \ -2,\ 0,\ 0,\ 0,\ 0,\ 0,\ 0)}$ & $1$ \\
\hline
\hline
$\{ X_1 \}\hspace{1cm} $ & $+$ & $3(3, 1)_{(1, \ 0, \ 0,\ 1, \ 0,\ 0,\ 0,\ 0,)}$ & $1/3$ \\
\hline
$\{ X_2 \}\hspace{1cm} $ & $+$ & $3(3, 1)_{(1, \ 0, \ 0,\ 0, \ -1,\ 0,\ 0,\ 0)}$ & $1/3$ \\
\hline
$\{ X_3 \}\hspace{1cm} $ & $+$ & $3(3, 1)_{(1, \ 0, \ 0,\ 0,\ 1,\ 0,\ 0,\ 0 )}$ & $1/3$ \\
\hline
$\{ X_4 \} \hspace{1cm}$ & $+$ & $3(3, 1)_{(1, \ 0, \ 0,\ 0,\ 0,\ -1,\ 0,\ 0 )}$ & $1/3$ \\
\hline
$\{ X_5 \}\hspace{1cm}$ & $+$ & $3(3, 1)_{(1, \ 0, \ 0,\ 0,\ 0,\ 1,\ 0,\ 0 )}$ & $1/3$ \\
\hline
$\{ X_6 \}\hspace{1cm} $ & $+$ & $3(3, 1)_{(1, \ 0, \ 0,\ 0,\ 0,\ 0,\ -1,\ 0 )}
$ & $1/3$ \\
\hline
$\{ X_7 \}\hspace{1cm} $ & $+$ & $3(3, 1)_{(1, \ 0, \ 0,\ 0,\ 0,\ 0,\ 1,\ 0 )}
$ & $1/3$ \\
\hline
$\{ X_8 \}\hspace{1cm} $ & $+$ & $3(3, 1)_{(1, \ 0, \ 0,\ 0,\ 0,\ 0,\ 0,\ -1 )}
$ & $1/3$ \\
\hline
$\{ X_8 \}\hspace{1cm}$ & $+$ & $3(3, 1)_{(1, \ 0, \ 0,\ 0,\ 0,\ 0,\ 0,\ 1 )}
$ & $1/3$ \\
\hline
\hline
$\{ H_d^1 \equiv L \} \hspace{1cm}$ & $+$ & $3(1, 2)_{(0, \ 1, \ 0,\ -1, \ 0,\ 0,\ 0,\ 0)}
$ & $-1/2$ \\
\hline
$\{ H_d^3 \} \hspace{1cm}$ & $+$ & $3(1, 2)_{(0, \ 1, \ 0,\ 0, \ -1,\ 0,\ 0,\ 0)}
$ & $-1/2$ \\
\hline
$\{ H_d^4 \} \hspace{1cm}$ & $+$ & $3(1, 2)_{(0, \ 1, \ 0,\ 0,\ 1,\ 0,\ 0,\ 0 )}
$ & $-1/2$ \\
\hline
$\{ H_d^5 \} \hspace{1cm}$ & $+$ & $3(1, 2)_{(0, \ 1, \ 0,\ 0,\ 0,\ -1,\ 0,\ 0 )}
$ & $-1/2$ \\
\hline
$\{ H_d^6 \} \hspace{1cm}$ & $+$ & $3(1, 2)_{(0, \ 1, \ 0,\ 0,\ 0,\ 1,\ 0,\ 0 )}
$ & $-1/2$ \\
\hline
$\{ H_d^7 \} \hspace{1cm}$ & $+$ & $3(1, 2)_{(0, \ 1, \ 0,\ 0,\ 0,\ 0,\ -1,\ 0 )}
$ & $-1/2$ \\
\hline
$\{ H_d^8 \} \hspace{1cm}$ & $+$ & $3(1, 2)_{(0, \ 1, \ 0,\ 0,\ 0,\ 0,\ 1,\ 0 )}
$ & $-1/2$ \\
\hline
$\{ H_d^9 \} \hspace{1cm}$ & $+$ & $3(1, 2)_{(0, \ 1, \ 0,\ 0,\ 0,\ 0,\ 0,\ -1 )}
$ & $-1/2$ \\
\hline
$\{ H_d^{10} \} \hspace{1cm}$ & $+$ & $3(1, 2)_{(0, \ 1, \ 0,\ 0,\ 0,\ 0,\ 0,\ 1 )}
$ & $-1/2$ \\
\hline
\hline
$\{ S_0 \}\hspace{1cm}$ & $+$ & $3(1, 1)_{(0, \ 0, \ -2,\ 0,\ 0,\ 0,\ 0,\ 0 )}$ & $0$ \\
\hline
$\{ S_1 \} \hspace{1cm}$ & $+$ & $3(1, 1)_{(0, \ 0, \ 1,\ -1,\ 0,\ 0,\ 0,\ 0 )}
$ & $0$ \\\hline
$\{ S_2 \equiv N_R \} \hspace{1cm}$ & $+$ & $3(1, 1)_{(0, \ 0, \ 1,\ 1, \ 0,\ 0,\ 0,\ 0)}
$ & $0$ \\
\hline
$\{ S_3 \} \hspace{1cm}$ & $+$ & $3(1, 1)_{(0, \ 0, \ 1,\ 0,\ -1,\ 0,\ 0,\ 0 )}
$ & $0$ \\
\hline
$\{ S_4 \} \hspace{1cm}$ & $+$ & $3(1, 1)_{(0, \ 0, \ 1,\ 0,\ 1,\ 0,\ 0,\ 0 )}
$ & $0$ \\
\hline
$\{ S_5 \} \hspace{1cm}$ &  $+$& $3(1, 1)_{(1, \ 0, \ 1,\ 0,\ 0,\ -1,\ 0,\ 0 )}
$ & $0$ \\
\hline
$\{ S_6 \} \hspace{1cm}$ & $+$ & $3(1, 1)_{(0, \ 0, \ 1,\ 0,\ 0,\ 1,\ 0,\ 0 )}
$ & $0$ \\
\hline
$\{ S_7 \} \hspace{1cm}$ & $+$ & $3(1, 1)_{(0, \ 0, \ 1,\ 0,\ 0,\ 0,\ -1,\ 0 )}
$ & $0$ \\
\hline
$\{ S_8 \} \hspace{1cm}$ & $+$ & $3(1, 1)_{(0, \ 0, \ 1,\ 0,\ 0,\ 0,\ 1,\ 0 )}$ & $0$ \\
\hline
$\{ S_9 \} \hspace{1cm}$ & $+$ & $3(1, 1)_{(0, \ 0, \ 1,\ 0,\ 0,\ 0,\ 0,\ -1 )}
$ & $0$ \\
\hline
$\{ S_{10} \} \hspace{1cm}$ & $+$ & $3(1, 1)_{(0, \ 0, \ 1,\ 0,\ 0,\ 0,\ 0,\ 1 )}
$ & $1/3$ \\
\hline\hline
$\{ X_{0} \} \hspace{1cm}$ & $+$ & $3({\bar 3}, 1)_{(-1, \ 0, \ -1,\ 0,\ 0,\ 0,\ 0,\ 0 )}
$ & $-1/3$ \\
\hline
\hline
$\{ Q_{1} \} \hspace{1cm}$ & $-$ & $3(3, 1)_{(2, \ 0, \ 0,\ 0,\ 0,\ 0,\ 0,\ 0 )}
$ & $2/3$ \\
\hline
$\{ Q_{2} \} \hspace{1cm}$ & $-$ & $3( 6^{*}, 1)_{(-2, \ 0, \ 0,\ 0,\ 0,\ 0,\ 0,\ 0 )}
$ & $-2/3$ \\
\hline
$\{ P_{1} \} \hspace{1cm}$ & $-$ & $3(1, 1)_{(0, \ 2, \ 0,\ 0,\ 0,\ 0,\ 0,\ 0 )}
$ & $-1$ \\
\hline
$\{ P_{2} \} \hspace{1cm}$ & $-$ & $3(1, 3^{*})_{(0, \ -2, \ 0,\ 0,\ 0,\ 0,\ 0,\ 0 )}
$ & $+1$ \\
\hline
\end{tabular}
\end{center}
\caption{\small
The chiral fermion spectrum of the 8-stack three generation N=1 MSSM on top of the table, in addition to three $N_R$'s and three pairs
of $H_u$, $H_d$ chiral matter. The model is non-supersymmetric but the MSSM matter is N=1 supersymmetric. Notice the presence of extra non-susy Higgsinos $H_d^k$, k=3, .., 10. 
\label{susymodels}}
\end{table}
The Yukawa couplings of the models are 
given \footnote{By ${\tilde H}_u$, ${\tilde H}_d^i$, ${\tilde S}_i$ 
we denote the boson tachyon superpartners of the chiral 
matter fields ${H}_u$, ${H}_d^i$, ${S}_i$. } 
by
\beq
Y \ = Y_{(SM)} + Y_{\mu} + Y_{X_0} + Y_{X_i},
\label{su1}
\eeq
where 
\beq
Y_{(SM)} = \ \lambda_{d} Q_L d_L^c {\tilde H}_d^2 + \lambda_{e}L e_L^{+} {\tilde H}_d^2 + 
\lambda_{\nu} L N_R {\tilde H}_u,
\label{lamix}
\eeq
\beqa
Y_{\mu}= \left( \lambda_{H_d} H_u H_d {\tilde S}_1 \right) + \{ \lambda_{H_d^1} L H_u {\tilde S}_2 \}+ 
+ \lambda_{H_d^3} 
H_u H_d^3 {\tilde S}_4 + \lambda_{H_d^4} H_u H_d^4 {\tilde S}_3 + 
\lambda_{H_d^5} H_u H_d^5 {\tilde S}_6  & \nonumber\\
+ \lambda_{H_d^6} H_u H_d^6 {\tilde S}_5 +  \lambda_{H_d^7} H_u H_d^7 {\tilde S}_8 
+  \lambda_{H_d^8} H_u H_d^8 {\tilde S}_7 + 
\lambda_{H_d^9} H_u H_d^9 {\tilde S}_{10} + \lambda_{H_d^{10}}  H_u H_d^{10} {\tilde S}_9 &\nonumber\\
\label{tropos}
\eeqa
\beqa
Y_{X_0}\ = \ \lambda_{N} X_0 d_L^c {\tilde N}_R \ + \lambda_{X_1^m} X_0  X_1 {\tilde S}_1 \ + \lambda_{X_2^m} X_0  X_2 {\tilde S}_4 \ +
\lambda_{X_3^m} X_0 X_3 {\tilde S}_3 \ + \lambda_{X_4^m} X_0  X_4{\tilde S}_6 \ +  \nonumber\\
\lambda_{X_5^m} X_0 X_5 {\tilde S}_5  + \lambda_{X_6^m} X_0 X_6 {\tilde S}_8  + 
\lambda_{X_7^m} X_0 X_7 {\tilde S}_7  +   \lambda_{X_8^m} X_0  X_8 {\tilde S}_{10}  +  
 \lambda_{X_9^m} X_0 X_9 {\tilde S}_9 \nonumber\\
\label{aytoi}
\eeqa
\beqa
Y_{X_i}\ = \ 
\lambda_{X_1} Q_L X_1 {\tilde H}_d^1 + 
\lambda_{X_2} Q_L X_2 {\tilde H}_d^4 + 
\lambda_{X_3} Q_L X_3 {\tilde H}_d^3 + 
\lambda_{X_4} Q_L X_4 {\tilde H}_d^6 \ + \ \nonumber\\ \lambda_{X_5}  Q_L X_5 {\tilde H}_d^5 \ 
+ \ \lambda_{X_6} 
Q_L X_6 {\tilde H}_d^8\  + \ \lambda_{X_7}  Q_L X_7 {\tilde H}_d^7\ + \nonumber\\
\ \lambda_{X_8}  Q_L X_8{\tilde H}_d^{10} 
\ + \ \lambda_{X_9}  Q_L X_9 {\tilde H}_d^9   
\label{upox1}
\eeqa
 $\bullet$ Quark, lepton and extra matter masses:

The first Yukawa term $Y_{(N=1 \ SM)}$ in (\ref{su1}) generates mass 
terms for all 
SM matter but the up-quarks, where we have identify the 
right neutrinos with the 
singlets $S_2$.
In the present models since a perturbative mass term the up-quarks is not allowed, it is interesting 
to engage in a short description of the phenomenology of these models.  
The ${\tilde H}_u$, ${\tilde H}_d^2$ Higgses - play 
the role of the corresponding two Higgs doublets
in the  
MSSM - couple to the SM matter at 
tree level. 
The neutrinos get also a tree level mass from a Yukawa term in (\ref{lamix}) as we have identify
$N_R \equiv S_2$.  
If for example we had identify $N_R$ with
e.g. $S_0$, the only mass term allowed for the neutrinos would have been the dimension eight operators that represent  $(\alpha^{\prime})^2$ corrections coming from the exchange of massive string states
\beq 
 m_{\nu}  \propto \frac{1}{M_s^4}L N_R \langle  Q_L d_L^c \rangle \langle ({\tilde S}_2\rangle)^2  \ .
\label{simi1}
\eeq
For values of the d-quark chiral condensate $\langle d_R d_L  \rangle \propto (200 \ MeV)^3 $  and values of the 
\beq
m_{\nu}  \propto \frac{< d_R d_L >}{M^2}
\label{simi}
\eeq
string scale $M_s \propto 1-10$ TeV neutrino masses of order 0.1 - 10 eV\rq{}s, consistent with neutrino oscillation experiments are obtained. Similar terms to (\ref{simi1}), that provide masses to neutrinos,  appear in the context of 4D IIA toroidal orientifold 5-, 6-models of 
\cite{kokos5}, \cite{kokos6}(see also \cite{louis2}) to be originating from the dimension six operators $(\alpha^{\prime}) L N_R \langle  Q_L u_L^c \rangle $. It appears that if the models had a low scale ${\cal O}(TeV)$ there would be a universality in the use of chiral condensate of QCD of giving  masses to neutrinos. However, in the present models the string scale may not be at the TeV region, as the branes wrap in all directions; there are no compact
directions transverse  to all stacks of branes \cite{D5}, \cite{savas}. 
\newline
$\bullet$ Colour triplets masses and bounds on $M_{X_i}$, $M_s$

An extended see-saw mass matrix is generated by the mixing - in 
(\ref{upox1}) and  (\ref{lamix})
-  between the d-quarks with the triplets $X_i$. 
If we neglect \footnote{The present mechanism of generating a Dirac mass of the triplets, is identical to the
one appearing in models of the fermionic formulation \cite{antonano}.}
the couplings (for simplicity) by assuming
that in (\ref{upox1}), (\ref{aytoi}) respectively
\beq
\lambda_{X_i} \rightarrow 0, i \neq 1; \ \   \lambda_{X_i^m} \rightarrow 0, \   \lambda_N  \rightarrow 0
\eeq
the see-saw generates
contributions to the scalar potential from  
the 1st term in (\ref{lamix}) and the 1st term in (\ref{upox1}), namely $\lambda_{d} Q_L d_L^c {\tilde H}_d^2 + \lambda_{X_1} Q_L X_1 {\tilde H}_d^1$
giving us mass eigenvalues for the quarks $(d_L^c, X_1 )$
of order 
\beq
m_d = \lambda_d \langle {\tilde H}_d^2 \rangle , \ 
m_{X_1} = \frac{ \lambda_{X_1}^2 }{ \lambda_d } \frac{ ({\langle {\tilde H}_d^1 \rangle})^2} { \langle {\tilde H}_d^2 \rangle} 
\label{mass}  
\eeq
In compactifications coming from intersecting branes, the couplings $\lambda_{X_1} \propto exp^{-A_{X_1} }$, ${\lambda_d}\propto \exp^{-A_d}  $ depend on the corresponding worldsheet areas $A_1$, $A_2$ connecting the intersection points taking part in the Yukawa interaction. Lets us select the maximum allowed value ${A_1}, A_2 \rightarrow 0$ , $\lambda_{X_1} \rightarrow 1 $, $\lambda_d  \rightarrow 1$. Then, as the mass of the \lq\lq{}heavy Higgs\rq\rq{} $\tilde H_d^1$, may be at least greater than 126 GeV (the mass of the LHC signal of \lq\lq{}lightest Higgs\rq\rq{}) 
and since $\lambda_d \langle {\tilde H}_d^2 \rangle = 4.8^{+0.7}_{-0.3} $ GeV \cite{oura},
we derive from (\ref{mass}) that the mass of the exotic heavy triplet $X_1$  should be \footnote{The current experimental limit in the appearance of a 
heavy quark is 100.8 GeV \cite{oura}.} at least :
\beqa
colour \ triplets \ bound : m_d = 5.5 \ MeV, \ m_{X_1} \geq 126 \ GeV \  \equiv m_{X_1} \geq 2.89 \times 10^6 \ GeV, &\nonumber\\
 colour \ triplets \ bound : m_d = 5.5 \ MeV, \ m_{X_1} \geq 124 \ GeV \  \equiv m_{X_1} \geq 2.79 \times 10^6 \ GeV,& \nonumber\\
\eeqa
 The latter puts a lower bound on the string scale which is \footnote{We are using the $\pm$ experimental errors in the mass of the d-quark. A value of $M_s \geq 3.53 \times 10^{6}$ Gev is obtained if we use the central value $m_d = 4.5$ MeV for $\langle {\tilde H}_d^2 \rangle = 126$ GeV.}
\beqa
M_{s} = 126 \ GeV, \ M_s \geq 2.89 \times 10^6 \ GeV\ ,   \nonumber\\
M_{s} = 124 \ GeV, \ M_s \geq 2.74 \times 10^6 \ GeV\  .   
\label{bound}
\eeqa
In principle, this bound can be made weaker if we allow for the exponentials in the Yukawa\rq{}s, to take non-zero values.  However, in this case, we may enter the area of 
$10^5 \ GeV = 10^2 \ TeV$ in which it has been argued \cite{abela1} that it is possible for flavour changing neutral current to appear in non-supersymmetric models from intersecting branes. At this stage of our understanding, it looks that the choice of maximum area is picked up only on anthropic principles.  
The value of new \lq\lq{}heavy quark\rq\rq{} $M_{X_1}$ mass is well within  the bounds predicted by ATLAS, where the search for a quark through the decay $X_1 \rightarrow  Z b$ excludes the new heavy down quark $X_1$ for masses below 725 GeV \cite{atlas1}.     
\newline
$\bullet$ higgsinos $\&$ $P_1, P_2, Q_1, Q_2$ masses

The first term in (\ref{tropos}) is the usual Dirac term for the Higgsinos.  The second term is a mixing term between the lepton L and the Higgsino $H_u$ generating mixing while the rest of the terms Dirac mass terms for the rest of Higgsinos.
Mass terms for the fields $P_1$, $P_2$ are given by the terms (respectively)  :
\beq
\frac{(P_1)^2 ({\tilde H}_u)^2 ({\tilde S}_1)^2 ({\tilde S}_2)^2}{M_s^6},  \ \frac{(P_2)^2 ({\tilde H}_d^3)^2 ({\tilde H}_d^4)^2}{M_s^3} \  ,
\label{terms}
\eeq
where ${\tilde S}_1$,  ${\tilde S}_2$ the scalar tachyonic superpartners of 
$S_1$, $S_2$ gauge fermion singlets.
Apart from the u-quark for which there is no obvious perrurbative mass term,
we also find that there are no mass terms for the $Q_1$, $Q_2$.


\subsection{A second example  of a non-susy model that accommodates the MSSM-matter using N=1 susy preserving D6-branes}

An alternative example of  a Standard-like model is obtained by 
changing the identification of fields that appear in table (\ref{susymodels}). 
For this purpose we identify the lepton field as $L \equiv H_d^3$. In 
table (\ref{tabmo1}), we list 
the chiral structure of the N=0 models. We have not included the 
$S_i$, $X_i$, $Q_i$, $P_i$ fields as they are the same as the ones appearing in 
table (\ref{susymodels}).
\begin{table}[htb] \footnotesize
\renewcommand{\arraystretch}{0.8}
\begin{center}
\begin{tabular}{|r|c|c|c|}
\hline\hline
 ${\bf Matter}\hspace{1cm}$ & $Massive$       &  ${(SU(3) \times SU(2))}_{(Q_a, Q_b, Q_{a1},Q_{c1}, 
Q_{c2}, Q_{c3},Q_{c4}, Q_{c5} )} \hspace{2cm}$ 
& $U(1)^{Y}$ \\
\hline
$ \{ Q_L \}\hspace{1cm} $ & $+$ & $3({3^{*}}, { 2^{*}})_{(-1,\ -1,\ 0,\ 0,\ 0,\ 0,\ 0,\ 0)} 
$ & $1/6$  \\
\hline
$\{ u_L^c  \}\hspace{1cm}$ & $-$ & $3(3, 1)_{(2,\ 0,\ 0,\ 0,\ 0,\ 0,\ 0,\ 0)}$ & $-2/3$  \\
\hline
$\{ d_L^c \}\hspace{1cm}$ &  $+$  & $3(3, 1)_{(1,\ 0, \ 0, \ -1,\ 0,\ 0, \ 0,\ 0)}$ & $1/3$ \\
\hline
$\{ H_u \}\hspace{1cm}$ & $+$ & $3(1, {2^{*}})_{(0,\  -1,\  -1,\ 0,\ 0,\ 0,\ 0,\ 0)}$ & $1/2$ \\
\hline
$\{ H_d^2 \}\hspace{1cm}$ & $+$ & $3(1, {2})_{(0,\ 1,\ 0,\ 1,\ 0,\ 0,\ 0,\ 0)}$ & $-1/2$ \\
\hline
$\{ e_L^{+} \}\hspace{1cm}$ & $+$ & $3(1, 1)_{(0, \ -2,\ 0,\ 0,\ 0,\ 0,\ 0,\ 0)}$ & $1$ \\
\hline
$\{ H_d^3 \equiv L \} \hspace{1cm}$ & $+$ & $3(1, 2)_{(0, \ 1, \ 0,\ 0, \ -1,\ 0,\ 0,\ 0)}
$ & $-1/2$ \\
\hline
$\{ S_2 \equiv N_R \} \hspace{1cm}$ & $+$ & $3(1, 1)_{(0, \ 0, \ 1,\ 1, \ 0,\ 0,\ 0,\ 0)}
$ & $0$ \\
\hline
\hline
$\{ H_d^1 \} \hspace{1cm}$ & $+$ & $3(1, 2)_{(0, \ 1, \ 0,\ -1, \ 0,\ 0,\ 0,\ 0)}
$ & $-1/2$ \\
\hline
$\{ H_d^4 \} \hspace{1cm}$ & $+$ & $3(1, 2)_{(0, \ 1, \ 0,\  0,\ 1,\ 0,\ 0,\ 0 )}
$ & $-1/2$ \\
\hline
$\{ H_d^5 \} \hspace{1cm}$ & $+$ & $3(1, 2)_{(0, \ 1, \ 0,\ 0,\ 0,\ -1,\ 0,\ 0 )}
$ & $-1/2$ \\
\hline
$\{ H_d^6 \} \hspace{1cm}$ & $+$ & $3(1, 2)_{(0, \ 1, \ 0,\ 0,\ 0,\ 1,\ 0,\ 0 )}
$ & $-1/2$ \\
\hline
$\{ H_d^7 \} \hspace{1cm}$ & $+$ & $3(1, 2)_{(0, \ 1, \ 0,\ 0,\ 0,\ 0,\ -1,\ 0 )}
$ & $-1/2$ \\
\hline
$\{ H_d^8 \} \hspace{1cm}$ & $+$ & $3(1, 2)_{(0, \ 1, \ 0,\ 0,\ 0,\ 0,\ 1,\ 0 )}
$ & $-1/2$ \\
\hline
$\{ H_d^9 \} \hspace{1cm}$ & $+$ & $3(1, 2)_{(0, \ 1, \ 0,\ 0,\ 0,\ 0,\ 0,\ -1 )}
$ & $-1/2$ \\
\hline
$\{ H_d^{10} \} \hspace{1cm}$ & $+$ & $3(1, 2)_{(0, \ 1, \ 0,\ 0,\ 0,\ 0,\ 0,\ 1 )}
$ & $-1/2$ \\
\hline
\hline
\end{tabular}
\end{center}
\caption{\small
A non-supersymmetric SM-like model with the three generation N=1 SM fermion spectrum on top
of the table, where only the 1 pair of exotics, namely $Q_i$, remain 
massless (as also $u_L^c$ does). 
Also present the $X_i$, $S_i$, $Q_i$, $P_i$ fermions seen in table (\ref{susymodels}). 
\label{tabmo1}}
\end{table}
Yukawa couplings for the quarks, leptons and the $X_i$'s are given by
\beq
{\tilde Y} \ = {\tilde Y}_{(N=1 \ SM)} + {\tilde Y}_{\mu} + 
{\tilde Y}_{X_0} + {\tilde Y}_{X_i},
\label{su2}
\eeq
where
\beq
{\tilde Y}_{N=1 \ SM} \ = \ \lambda_d Q_L d^c_L {\tilde H}_d^2 \ + \  \lambda_e L e_L^{+} {\tilde H}_u^4 + 
 \lambda_{\nu} \frac{L N_R {\tilde H}_u {\tilde S}_o {\tilde S}_1 {\tilde S}_4}{M_s^3}
\eeq

\beq
{\tilde Y}_{\mu} = Y_{\mu}, \  \ {\tilde Y}_{X_o} = Y_{X_o}, \ {\tilde Y}_{X_i} = Y_{X_i}, \ 
\eeq
As in the models of the previous section, the non-chiral fields $Q_i$ $\&$ $u_L^c$ remain massless. 
Other N=0 SM-like models can be obtained from the one\rq{}s in table (\ref{susymodels})  by changing the assignment of  right 
handed neutrinos to any of the singlets $S_i$ and/or the identification of leptons with any of the fields $H^i_d$.


\section{IIA $Z_3 \times Z_3$ orientifolds on the ABB lattice} 

$\bullet$ {\bf Spectrum rules}

For the calculations on the ABB lattice in this section, we may use a different lattice basis that the one used in the rest of this 
work. We are using the A, B lattices found in the Appendix of \cite{koz3z3} and in \cite{lust3}, where 
a general D6$_a$-brane is determined by three pairs of wrapping numbers
$(n_a ,m_a )$ along the fundamental cycles (organized into orbits) with complex structure in each $(T^2)_a$-torus  a=1,2,3, defined as $U_A =\frac{1}{2}+ i \frac{\sqrt{3}}{2}$, $U_B = \frac{1}{2}+i \frac{1}{2 \sqrt{3}}$.
A generic D6$_a$-brane is determined by three pairs of wrapping numbers
$(n_a ,m_a )$ along the fundamental cycles of each $T^2$, 
\beq
e^A_1 = e^B_1 = R, \ e^A_2 =\frac{R}{2}+ i \frac{\sqrt{3R}}{2}, \ e^B_2 =\frac{R}{2}+ i\frac{R}{2\sqrt{3}} 
\  . 
\eeq
The 
massless spectrum is given in terms of the effective wrappings $({\tilde Z}_a, {\tilde Y}_a)$. 
The RR tadpoles, the cancellation of RR charge in homology, are found to be   
\beq
\sum_a N_a {\tilde Z}_a = 12,
\label{tadp}
\eeq
\beq
{\tilde Z}_a \ = \ 6 n^1_a n^2_a n^3_a  + 3 (n^1_a n^2_a m^3_a + n^1_a m^2_a n^3_a  + m^1_a n^2_a n^3_a) + m^2_a m^3_a n^1_a - m^1_a m^2_a m^3_a
\eeq
or
\beq
{\tilde Z}_a \ = \ 6 n^1_a n^2_a n^3_a  + 3 (\underline{n^1_a n^2_a m^3_a}) + m^2_a m^3_a n^1_a - m^1_a m^2_a m^3_a \ ,
\eeq
where by underline we mean all possible permutations of indices and 
\beq
{\tilde Y}_a \  = \  3 n_a^1 n_a^2 n_a^3  + n_a^1 n_a^2 m_a^3 + n^1_a m^2_a n^3_a - m_a^1 n_a^2 m_a^3 - m_a^1 m_a^2 n_a^3 - 
m_a^1 m_a^2 m_a^3 \ .
\eeq
\begin{table}[htb] \footnotesize
\renewcommand{\arraystretch}{2}
\begin{center}
\begin{tabular}{|c|c|}
 \hline
 ${\bf Representation\hspace{1cm}}$ & $ {\bf Multiplicity \hspace{1cm}}$ \\\hline
 $\hspace{2cm} \Yasymm_L \ = \ A_L  \hspace{1cm}$   & $ \frac{1}{2}(I_{aa^{\prime}} + I_{a, O6}) \ = \ 
\frac{3}{2} (Z_a - 2Y_a)(Z_a + 3)     \hspace{1cm}$ \\
\hline
  $\hspace{2cm}   \Ysymm_L \ = \ S_L  \hspace{1cm}$   & $ \frac{1}{2}(I_{aa^{\prime}} - I_{a, O6}) \ = \ 
\frac{3}{2} (Z_a - 2Y_a)(Z_a - 3)  \hspace{1cm}$  \\
 \hline
  $\hspace{2cm} (N_a, {\bar N}_b)_L        \hspace{1cm}$ & $ I_{ab} \ = \ 3 (Z_a Y_b - Y_a Z_b)    \hspace{1cm}$ \\
\hline   
  $\hspace{2cm}     (N_a,  N_b)_L             \hspace{1cm}$ & $ I_{ab^{\prime}} \ = \   3 (Z_a Z_b -Z_a Y_b - Y_a Z_b)    \hspace{1cm}$ \\
\hline\hline
\end{tabular}
\end{center}
\caption{\small The massless open string spectrum in four dimensions (ABB lattice).
\label{abblat}}
\end{table}
Table (\ref{abblat}) contains the usual left handed bifundamental fields $(N_a, {\bar N}_b)$, $(N_a, {N}_b)$, as well 
as antisymmetric and symmetric representations of chiral open strings stretching between D6-branes and its nine images 
$\Omega R \Theta^k \omega^l$.
Also present are massless adjoint, non-chiral, matter created by open strings stretching between the D6-branes and 
its $ \Theta^k \omega^l$ image branes as follows :
\beqa
(Adj)_L \ : \ [ (m^1)_a^2 + (n^1)_a^2 +(m^1)_a (n^1)_a   ][(m^2_a m^3_a)^2 \ + \ 3 m^2_a (m^3_a)^2 n^2 \ 
+ \ 3 (m^3_a)^2 (n^2_a)^2 \ + \nonumber\\ 3 (m^2_a)^2 m_a^3 n_a^3  + 3 (m^2_a)^2 (n^3_a)^2 + 
 9 m^2_a m^3_a n^2_a n^3_a  +  9 m^3_a (n^2_a)^2 n^3_a + 9 m_a^2 n_a^2 (n^3_a)^2 + 9 (n^2_a)^2 (n^3_a)^2 ] \nonumber\\
\eeqa
Adjoint matter is N=1 supersymmetric as the $Z_3 \times Z_3$ rotations preserve N=1 supersymmetry. 

$\bullet$ {\bf Anomaly cancellation}

The cancellation of cubic non-abelian gauge anomaly is proportional to
\beq
\sum_{b \neq a}(N_b Z_a Z_b - 2 N_b N_b Z_b Y_a) + (N_a - 4)(Z_a -2 Y_a)(Z_a + 3)+(N_a + 4)(Z_a - 2 Y_a)(Z_a - 3)
\eeq 
The anomaly is cancelled through the use of tadpole condition (\ref{tadp}).
The mixed non-abelian U(1) - $SU(N_b)^2$ anomalies may read 
\beq
A_{mixed} = \frac{N_a}{2} (I_{ab}+I_{ab^{\prime}} ) = \frac{N_a ({\tilde Z}_a - 2 {\tilde Y}_a){\tilde Z}_b}{2}
\label{mixed}
\eeq

In order to cancel these anomalies one has to make use of a generalized Green-Schwarz mechanism (GSM) as suggested in 
\cite{fra},  \cite{louis2}. The GSM application in the case of of our $Z_3 \times Z_3$ orientifolds reads
\beqa
\sum_a N_a ( M_o - (\Omega R) M_o ) \int_{M4} B^0_2 \wedge F_a \ \ ;& \ \  (K_o + (\Omega R)K_o)\int_{M4} C^0 \wedge F_b \wedge F_b\nonumber\\
\sum_a N_a  ( N_o - (\Omega R) N_o ) \int_{M4} B^I_2 \wedge F_a \ \ ;& (L_o +(\Omega R) L_o )^I \int_{M4} C^I \wedge F_b  \wedge F_b .
\label{GSC}
\eeqa
\beqa
M_o = \sum_{orbits} m^1_a m^2_a m^3_a, &K_o&= \sum_{orbits} n^1_b n^2_b n^3_b \nonumber\\
N_o = \sum_{orbits} n^J_a n^K_a m^I_a &L_o&= \sum_{orbits} n^I_b m^J_b m^K_b
\eeqa
 \beqa 
(M_o - (\Omega R) M_o) = -18({\tilde Z}_a -2 {\tilde Y}_a), &&(K_o + (\Omega R)K_o)  = -3 {\tilde Z}_b ,\nonumber\\
(N_o - (\Omega R) N_o)^1 = -3({\tilde Z}_a -2 {\tilde Y}_a) , && (L_o +(\Omega R) L_o)^1 = - 9 {\tilde Z}_b  ,\nonumber\\
(N_o - (\Omega R) N_o)^2 = 0 , && (L_o +(\Omega R) L_o)^2 = -3 {\tilde Z}_b  , \nonumber\\
(N_o - (\Omega R) N_o)^3 = 0 , && (L_o +(\Omega R) L_o)^3 = -3 {\tilde Z}_b
\eeqa

It is obvious, that the couplings (\ref{GSC}) have the right form to cancel the anomaly (\ref{mixed}).  
\newline
\newline
$\bullet$ {\bf Model Building on the ABB lattice}
\newline

The choise of wrappings 
\beqa
  (Z, Y)_{a}& = & (3, 1), \  \ (Z, Y)_{b} = (1, 0), \  \ (Z, Y)_{c} = (1, 1),    
\label{abbla}
\eeqa
satisfies the RR tadpoles (\ref{tadp}). The associated non-susy chiral spectrum
is seen on table (\ref{ffd}).

\begin{table}[htb] \footnotesize
\renewcommand{\arraystretch}{1.5}
\begin{center}
\begin{tabular}{|r|c|c|}
\hline\hline
 ${\bf Matter\hspace{2cm}}$ & $(SU(3) \times SU(2))_{(Q_a, Q_b, Q_c)} 
\hspace{2cm}$ & $U(1)^{Y}$ \\
\hline\hline
$\hspace{2cm} \{ 6 {\bar Q}_L \} \hspace{2cm}$   & $3({\bar 3}, {2}, 1)_{(-1,\ 1,\ 0)} \hspace{2cm}$ & $-1/6$  \\
\hline
$\hspace{2cm} \{ 3 Q_L \} \hspace{2cm}$   & $6({3}, {2}, 1)_{(1,\ 1,\ 0)} \hspace{2cm}$ & $1/6$  \\
\hline
$\hspace{2cm}\{ 6 X_1 + 3 d_L^c \}\hspace{2cm}$   & $9({3}, 1, 1)_{(2,\ 0,\ 0)}\hspace{2cm}$ & $1/3$  \\
\hline
$\hspace{2cm}\{ 6 X_2  \}\hspace{2cm}$ & $6(3, 1, 1)_{(1,\ 0, \  -1)}\hspace{2cm}$ & $-1/3$ \\
\hline
$\hspace{2cm}\{ 3 u_c^L \}\hspace{2cm}$ & $3({\bar 3}, 1, 1)_{(-1,\ 0, \  -1)}\hspace{2cm}$ & $-2/3$ \\
\hline
$\hspace{2cm}\{ 3 L \}\hspace{2cm}$ & $3(1, 2, 1)_{(0,\  1,\  -1)}\hspace{2cm}$ & $-1/2$ \\
\hline
$\hspace{2cm}\{ 3 S_1 + 3 N_R \}\hspace{2cm}$ & $6(1, 1, 1)_{(0,\  2,\  0)}\hspace{2cm}$ & $0$ \\
\hline
$\hspace{2cm}\{ 3 S_2 \}\hspace{2cm}$ & $3(1, {\bar 3}, 1)_{(0, \ -2,\ 0)}\hspace{2cm}$ & $0$ \\
\hline
$\hspace{2cm}\{ e_R^c \}\hspace{2cm}$ & $3(1, 1, 1)_{(0, \ 0, \ 2)}\hspace{2cm}$ & $1$ \\
\hline
\end{tabular}
\end{center}
\caption{\small
A three generation non-supersymmetric chiral (open string) spectrum accommodating the MSSM fermionic matter. The 
 hypercharge is $U(1)^Y = \frac{1}{3}F_a +\frac{1}{2}F_c$.    
\label{ffd}}
\end{table}


\section{ SPLIT SUPERSYMMETRY $\&$ D6-brane 
models with $sin^2( \theta) =\frac{3}{8}$ }

String Theory Intersecting D-brane models (STIB) is the natural arena for the realization of 
ideas on the 
existence of split supersymmetry (SS) \cite{split1, split2, split3} 
in particle physics. 
The SS claim \cite{split1} relies on a number of assumptions that demand :\newline
(a) that the particle spectrum of the SM remain massless to low energies \newline
(b) that the SM spartners become massive with a mass of the order of the supersymmetry 
breaking scale \newline
(c) the gauge couplings unify at a scale near $10^{16}$ GeV \newline
(d)   there are light gauginos in the presence of gravity [Note that condition 
should be modified in STIBs as gauginos get a mass of order $M_s$.] \newline 
(e) there are light higgsinos in the presence of gravity and thus might be 
seen experimentally\newline 
(f) the assumption of the existence of a heavy 
 and a light Higgs set of doublets present in the spectrum that form two 
chiral supermultiplets.
In this section we will present models that can some times fully satisfy conditions (a), 
(b), (c) (d), (e).

We also note that condition (f), the existence of a 
 Higgs light was assumed in \cite{split1} that it may be a result of a fine tuning
mechanism. In models that may come from intersecting branes supersymmetric Higgs sets may 
appear naturally. 
The present D-brane models have not a supersymmetric Higgs sector. However, the Higgs system
can be understood as part for the massive 
spectrum that organizes itself in terms of massive N=2 hypermultiplets. 
The Higgs fields become subsequently tachyonic
in order to participate in electroweak symmetry breaking [see also similar considerations 
 \cite{louis2,  
kokos5, kokos6, kokos1}].

Conditions (a), (b), (c), (d) can be naturally obtained - in intersecting brane worlds - 
and in the current models. The (e), (f) conditions are harder to be 
obtained and may be examined in a case by case basis. Also condition (c) for a 
particle physics model that includes gravity - as STIBs - means that 
the string scale should be high and at least  $10^{16}$ GeV. 
Condition (a) can be naturally satisfied in STIBS and there are a lot of models 
exhibiting only the SM spectrum at low energies. These models 
accommodate the SM spectrum and
can be either belong to an overall non-supersymmetric 
model or to an overall N=1 supersymmetric model.
The first case involves three generation models  
with supersymmetry broken
at the string scale as it has been exhibited in the toroidal orientifolds
in \cite{louis2, kokos5, kokos6, kokos1} and in the previous sections as
superpartners of the SM are being massive and of the order of the string scale.  
It is also possible to construct non-supersymmetric constructions that localize locally
the N=1 supersymmetric SM spectrum and such models have been 
constructed in \cite{cre1}, \cite{kokosusy}, \cite{floko}. 
In particular in \cite{kokosusy} we generalized the four stack constructions 
of \cite{cre1} by 
also including a non-zero B-field flux in the models[also extending these models
to their maximal extensions with gauge groups made of 
five and six stacks of D6-branes at the string scale]. This makes the torus tilted and 
allows for more general solutions in the RR tadpoles as well changing the number
of N=1 Higgs supermultiplets present in the spectrum.  
In the four stack models constructed in \cite{cre1, kokosusy} it has been 
shown \cite{gauge} that it is possible to accommodate the successful prediction of 
supersymmetric SU(5) GUTS with $sin^2(\theta) = 3/8 $ at a string scale which coincides
with the unification scale of $2 \cdot 10^{16}$ GeV, and 
all gauge coupling constants unified at $10^{16}$ GeV. Hence condition (c) is also
satisfied in IBs and obviously all the D-brane models appearing in 
 \cite{cre1}, \cite{kokosusy} could form realistic 
D-brane split susy models as RR tadpoles has been shown to be consistently 
implemented in \cite{floko}and in the present work. 
Next, we will also show that it is also possible - in the framework of the 
present $Z_3 \times Z_3$ orientifolds - to easily 
build models which 
satisfy conditions (a),(b),(c),(d); some models may also satisfy partially or 
fully the condition (e).


\subsection{1st example : Split Susy (SM + $\nu_R$ + Higgsinos) models: \lq\lq{}massive Gauginos\rq\rq{} $\&$ \lq\lq{}light  Higgsinos\rq\rq{} $\&$ $sin^2 \theta_W(M_S)  = \frac{3}{8}$ }

In this section we will construct a deformation of the SM's appeared in section (3.1) and table 
(\ref{taboldzero}) where (a), (b), (c), (d), (e) conditions of split susy are satisfied.  
These models have the initial gauge group $U(3) \times U(2) \times U(1)$ with supersymmetry broken at $M_s$ and at low energy 
the gauge group becomes identical to the SM. The massless fermion spectrum contains the MSSM fermion matter and 
is given in table (\ref{give1}). 
RR tadpoles are satisfied by the choices $N_a = 3$, $N_b = 2 $, $N_c = 1$ and  
\beq
(Z_a, Y_a) = \left( \ba{cc} 1,  & 1 \\\ea \right), \ (Z_b, Y_b) = 
\left( \ba{cc} 1, & 1 \\\ea \right), \ (Z_c, Y_c) = \left( \ba{cc}-1, & 1 \\\ea \right)  \ . 
\label{wrap3}
\eeq 
At the top of the table (\ref{give1}) we see the massless fermion spectrum of the 
N=1 SM whose corresponding superpartners are part of the massive spectrum and appear in the 
 intersection of each corresponding fermion; hence condition (b) is satisfied.
\begin{table}[htb] \footnotesize
\renewcommand{\arraystretch}{1.8}
\begin{center}
\begin{tabular}{|r|c|c|c|}
\hline\hline
 ${\bf Matter\hspace{2cm}}$ & Intersection & $(SU(3) \times SU(2))_{(Q_a, Q_b, Q_c)} \hspace{2cm}$ & $U(1)^{Y}$ \\
\hline\hline
$\hspace{2cm} \{ Q_L \} \hspace{2cm}$   & $ab*$ & $3({3}, { 2})_{(-1,\ -1,\ 0)} \hspace{2cm}$ & $1/6$  \\
\hline
$\hspace{2cm}\{ u_L^c  \}\hspace{2cm}$   & $A_a$   & $3(3, 1)_{(-2,\ 0,\ 0)}\hspace{2cm}$ & $-2/3$  \\
\hline
$\hspace{2cm}\{ d_L^c \}\hspace{2cm}$ &  $ac$   & $3(3, 1)_{(1,\ 0, \  -1)}\hspace{2cm}$ & $1/3$ \\
\hline
$\hspace{2cm}\{ \ L \}\hspace{2cm}$ & $bc$  & $3(1, 2)_{(0,\  1,\  -1 )}\hspace{2cm}$ & $-1/2$ \\
\hline
$\hspace{2cm}\{ \ H_d \}\hspace{2cm}$ &bc$ $  & $3(1, 2)_{(0,\  1,\  -1 )}\hspace{2cm}$ & $-1/2$ \\
\hline
$\hspace{2cm}\{ H_u \}\hspace{2cm}$ & $bc*$   & $3(1, {\bar 2})_{(0,\  -1,\  -1 )}\hspace{2cm}$ & $1/2$ \\
\hline
$\hspace{2cm}\{ e_L^{+} \}\hspace{2cm}$ & $A_b $  & $3(1, 1)_{(0, \ -2,\ 0)}\hspace{2cm}$ & $1$ \\
\hline
$\hspace{2cm}\{ N_R \}\hspace{2cm}$ & $S_c $  & $9(1, 1)_{(0, \ 0, \ 2)}\hspace{2cm}$ & $0$ \\
\hline
\hline
$\hspace{2cm}\{ C_1 \}\hspace{2cm}$ & $ac $  & $3(3, 1)_{(1, \ 0, \ -1 )}\hspace{2cm}$ & $1/3$ \\
\hline 
$\hspace{2cm}\{ C_2 \}\hspace{2cm}$ & $ac* $  & $3({\bar 3}, 1)_{(-1, \ 0, \ -1 )}\hspace{2cm}$ & $-1/3$ \\
\hline
\hline
\end{tabular}
\end{center}
\caption{\small
A three generation 4D non-supersymmetric model with the chiral content of 
N=1 MSSM on top of the table, in addition to $N_R$'s. There are three pairs
of $H_u$, $H_d$ Higgsinos. This model
mimics models coming from gauge mediation scenarios and possess 
$sin^2 \theta = 3/8$ at $M_s$.  
\label{give1}}
\end{table}
In order to show that at low energy only the SM remains [condition (a)]; the extra beyond the SM fermion matter and all U(1) gauge fields originally present at the string scale should become 
massive but the hypercharge.  Hence Higginos and the exotics 
$C_i$ receive non-zero masses. The Yukawa couplings (\ref{yuk}) provide masses to d,e and a Dirac term for neutrinos
\beq
\lambda_d {(Q_L)}_{(-1, -1, 0)} {(d^c_L)}_{(1, 0, -1)} \langle {{\tilde H}_d} \rangle \ + \ \lambda_{\nu} L N_R \langle {\tilde H}_u \rangle \  + \frac{\lambda_e}{M_s}L e_L^c \langle {\tilde H}_d \rangle   + (\lambda_H H_u H_d + \lambda_C C_1 C_2) \langle {\tilde N}_R \rangle
\label{yuk}
\eeq
via the previously massive tachyonic superpartners ${\tilde H}_u$, ${\tilde H}_d$ of  
Higgsinos $H_u$, $H_d$ as in eq. (\ref{choices}).  
We also note that the Higgs fields have the quantum numbers
\beq
{\tilde H}_u = (1, 2, 1)_{(0, -1, -1)}, \ \ {\tilde H}_d = (1, 2, 1)_{(0, 1, 1)} \ . 
\eeq
The pair of non-chiral colour 
triplets could get massive by their Yukawa coupling to the tachyonic spartner of $N_R$, $ \langle {\tilde N}_R \rangle$
(by choosing $\lambda_C \sim e^{-A_{C}}, \ A_C \rightarrow 0$) of order $M_s$.  
The Higgsinos [condition (e)] $H_u$, $H_d$, form a Dirac mass term from the Yukawa coupling 
\beq
\lambda_H H_u H_d \langle {\tilde N}_R \rangle , \  \ \lambda_H  \sim e^{- A_{hig}} \ , 
\label{higs}
\eeq
 where  ${\tilde N}_R$ is the scalar tachyonic superpartner 
of $N_R$. The vev of ${\tilde N}_R$ is of the order of the string scale when $A_{hig} \rightarrow 0$. Due to nature of the
Yukawa coupling term [see also eqn. (\ref{supr1})] in (\ref{higs}) the Higgsino mass 
can be anywhere between the string scale and the scale of electroweak symmetry breaking, 
\beq
100 \  GeV < m_{(Higgino \ pair)} \leq M_s \ , 
\label{maxhigg1}
\eeq
which requires the areas to be $A_{min} \sim 34 $, $A_{max} = 0$ for the lower and upper 
limits of (\ref{maxhigg1}).
Lets us discuss the U(1) structure.
One U(1) gauge field 
becomes massive through its BF couplings, 
namely $- 3F_a - 2F_b -3F_c$, while from the extra U(1)'s that survive 
massless the GS mechanism; one combination of U(1)'s is the hypercharge while the third  
$U(1)^{add}$ gets broken by the vev of ${\tilde N}_R$.
\beq
 U(1)^Y = \frac{1}{3} F_a - \frac{1}{2}F_b, \  \  
U(1)^{add}= F_a + (2/3)F_b - (13/9)F_c \ .  
\eeq
As the present $Z_3 \times Z_3$ orientifolds involve D6-branes 
the gauge coupling constants are controlled by the length of the corresponding cycles 
that the D6-branes wrap 
\beq
\frac{1}{\alpha_a} \ = \ ||l_i|| \ ,
\eeq  
where $||l_i||$ is the length
of the corresponding cycle for the i-th set of brane stacks. 
The canonically normalized U(1)'s as well the normalization of the abelian generators are given by
\beq
{\tilde U}(1)_a \ = \  \frac{F_a}{\sqrt{2 N_a}}, \ \ Tr(T_a T_b) = \frac{1}{2} \delta_{ab}
\eeq
The hypercharge is given \footnote{we used the conventions used in \cite{toma}} as a linear combination $Y(1)^Y = \sum_i c_i F_i$; hence in the present models
the
value of the weak angle is computed to be     
\beq
\sin^2 \theta_W = \frac{1}{1 + 4c_2^2 + 6c_3^2 (\alpha_2/\alpha_3)}  \ .
\eeq
Taking into account that in the present \footnote{The $(Z_i, Y_i)$ ,  i = a, b and subsequently $ (n_i, m_i)$ are identical.  
} models $\alpha_2 = \alpha_3$, the gauge couplings unify at $M_s$,  we get 
\beq
\sin^2 \theta_W \stackrel{M_s}{=} \frac{3}{8} \  .
\label{sin}
\eeq
Gauginos are massless at tree level in the present intersecting brane models 
and appear in the
four dimensional N=4 SYM spectrum that get localized from strings having both ends on the same
set of D6-branes. A mechanism for generating gaugino masses in intersecting branes - 
due to quantum corrections - have been 
put forward in \cite{louis2} where non-supersymmetric toroidal orientifold models are discussed 
\cite{louis2, kokos5, kokos6, kokos1}. 
 According to this result \cite{louis2}, as gauginos are massless at tree level, 
loop corrections to gauginos proceed
via massive fermions running in the loops. The order of gaugino masses 
is of the order of the supersymmetry breaking scale, the string scale.
The same mechanism may persist in the present models.  
A comment is in order. The spectrum of table (\ref{give1}) is invariant under the 
exchanges $C_1 \leftrightarrow d_L^c$, $L \leftrightarrow  H_d$.

The D6-brane non-susy MSSM-like quiver configuration of table (\ref{give1})   
have been simultaneously suggested in \cite{noti1} (without the triplets) in a local model context, as a string theory realization of split supersymmetry (see note added in the end of this work). For further 
studies of spit supersymmetry see \cite{split2, split3}. 

\subsection{2nd example :  Split Susy (SM + $\nu_R$) models: \lq\lq{}massive\rq\rq{} Gauginos,  \lq\lq{}massive Higgsinos\rq\rq{}, $sin^2 \theta_W (M_s) = 3/8$ }

In this section, we will present models that satisfy (a),(b), (c), (d) conditions of split 
susy.
These models are build from three stacks of intersecting 
D6-branes at the string scale; a variant of the models considered in section (4.1). 
The satisfaction of the RR tadpoles (\ref{tad}) proceeds via the choice
\beq
N_a \ = \ 3, \ \ N_b \ = \ 2, \  \ N_c \ = \ 1   
\eeq
and with the choice of effective wrappings
\beq
(Z_a, Y_a) = \left( \ba{cc} 1,  & 0 \\\ea \right), \ (Z_b, Y_b) = 
\left( \ba{cc} 1, & 0 \\\ea \right), \ (Z_c, Y_c) = \left( \ba{cc}-1, & 0 \\\ea \right)
\label{len1}
\eeq 
The SM fermion spectrum with three quark and lepton families is seen in table (\ref{tabse1}).
\begin{table}[htb] \footnotesize
\renewcommand{\arraystretch}{2}
\begin{center}
\begin{tabular}{|r|c|c|}
\hline\hline
 ${\bf Matter\hspace{2cm}}$ & $(SU(3) \times SU(2))_{(Q_a, Q_b, Q_c)} \hspace{2cm}$ & $U(1)^{Y}$ \\
\hline\hline
$\hspace{2cm} \{ Q_L \} \hspace{2cm}$   & $3(3, {2})_{(1,\ 1,\ 0)} \hspace{2cm}$ & $1/6$    \\
\hline
$\hspace{2cm}\{ u_L^c \}\hspace{2cm}$   & $3({\bar 3}, 1)_{(2,\ 0,\ 0)}\hspace{2cm}$ & $-2/3$   \\
\hline
$\hspace{2cm}\{ d_L^c \}\hspace{2cm}$ & $3(3, 1)_{(-1,\ 0, \  1)}\hspace{2cm}$ & 
$1/3$ \\
\hline
$\hspace{2cm}\{ L \}\hspace{2cm}$ & $3(1, 2)_{(0,\  -1,\  -1 )}\hspace{2cm}$ & 
$-1/2$  \\
\hline
$\hspace{2cm}\{ e_L^{+} \}\hspace{2cm}$ & $3(1, 1)_{(0, \ -2,\ 0)}\hspace{2cm}$ & 
$1$  \\
\hline
$\hspace{2cm}\{ N_R \}\hspace{2cm}$ & $3(1, 1)_{(0, \ 0, \ 2 )}\hspace{2cm}$ & $0$  \\
\hline
\end{tabular}
\end{center}
\caption{\small
A three generation chiral (open string) spectrum accommodating the SM. 
The required Higgs may come from 
 bifundamental N=2 hypermultiplets in the N=2 $bc$, $bc^{\star}$ 
sectors \cite{louis2, kokos5, kokos6} that may trigger brane recombination.
\label{tabse1}}
\end{table}
There is one anomalous U(1) which becomes massive
\beq
U(1)^{massive} = 3 F_a + 2 F_b -2 F_c
\eeq 
and two anomaly free U(1)'s. One of them which can be identified
\beq
U(1)_{ex}  \ = \ 3 F_a + 2 F_b -\frac{13}{2} F_c
\eeq
becomes massive by the vev of the tachyonic scalar superpartner of the right handed neutrino, 
leaving only the hypercharge massless to low energies 
\beqa
U(1)^Y = -\frac{1}{3}F_a  + \frac{1}{2}F_b   \  . 
\label{free}
\eeqa

Light Higgsinos are not present in the models considered in this section as 
they are part of 
the massive spectrum with a mass of the order of the string scale ( having chosen $A=0$; see (\ref{maxhigg1})); in general massive with 
a mass above the electroweak scale. 
Gauginos are expected to receive string scale masses. The strong and weak gauge couplings
also unify as here also $a_2 =a_3$ and $sin^2(M_{GUT}) = 3/8$.


\section{Non-supersymmetric gauge unification for intersecting brane split susy models}

Gauge unification for supersymmetric intersecting brane models has been examined in \cite{gauge}.  
The evolution of the one loop renormalization group equations (ERGE) for the $SU(3)_c$, $SU(2)_w$, $U(1)_Y$ gauge couplings in the absence of one-loop string threshold corrections  
(see \cite{lsti} and also \cite{kokothre}), $a_2 (M_Z) = a_{em} sin^2 \theta(M_Z)$,  is given by 
\beqa
\frac{1}{a_s (M_z)} = \frac{1}{a_s (M_s)} - \frac{b_3}{2 \pi} \ln \left( \frac{M_Z}{M_s}\right), \nonumber\\
\frac{sin^2 \theta_w (M_Z)}{a_{em} (M_Z)} = \frac{1}{a_w (M_s)} -\frac{b_2}{2\pi} \ln \left(\frac{M_Z}{M_s}\right),\nonumber\\
\frac{cos^2\theta_w (M_Z)}{a_{em} (M_Z)} = \frac{1}{a_Y (M_s)} -\frac{b_1}{2\pi} \ln \left( \frac{M_Z}{M_s}\right),
\label{RG}
\eeqa
\beq
\frac{2}{3} \frac{1}{ \alpha_s (M_Z)}  + \frac{2 sin^2 \theta_w (M_Z) - 1}{\alpha_{em} (M_Z)} =\frac{B}{2 \pi} \ \ln \frac{M_Z}{ M_s} \ ,
\label{strings}
\eeq
and $B=-(\frac{2}{3} b_3 + b_2 - b_Y)$; $b_3$, $b_2$, $b_1$ are the $\beta$-function coefficients for strong, 
weak and hypercharge gauge couplings. For a theory which accommodates the SM and a number of extra particles below a scale $M_s$
\beq
 ( b_1, \ b_2 , \ b_3 ) \ = \ ( \frac{20}{9}n_G + \frac{1}{6}n_H + N_1,\   \frac{4}{3}n_G + \frac{1}{6}n_H + N_2 -\frac{22}{3},   \ 
\frac{4}{3}n_G -11 + N_3),  
\label{betas}
\eeq
where $N_1, N_2, N_3$ the contribution of the beyond the SM particles; the rest of the terms in 
(\ref{betas}) are the Standard model contributions; $n_G$ the number of generations; $n_H$ 
the number of Higgses. 
Lets us now examine gauge coupling unification using the MSSM-matter like models of table (\ref{give1}) as a representative example. They are non-supersymmetric but they respect $sin^2 (\theta_W )= 3/8$ at the unification scale, as $a_2 = a_3$. This implies that at the 
unification scale where the gauge couplings $SU(2)_w$, $SU(3)_c$ meet, we have
the standard SU(5) relation 
\beq
(5/3) a_Y = a_s = a_w = a_2 = a_3 
\eeq
This relation \lq{}solves\rq{} the following eqn.
\beq
\frac{1}{a_Y}=\frac{2}{3}\frac{1}{a_2} +\frac{1}{a_w} \ .
\eeq
Remember that in traditional GUTS one has to normalize under SU(5) the RG equations, to achieve unification. 
On the contrary, take a a representative example the models of table (\ref{give1}), supersymmetry is already broken at the string scale 
and the theory is non-supersymmetric all along to low energies of electroweak order. 
The full 
massless spectrum of the model  contains, the usual SM and nine right handed neutrinos (all with the same U(1) 
charges), in addition to three \footnote{Models with (multiplicity) 3 pairs of up, down higgsinos have also appeared in section 4.1 in the 
non-susy 4D models of \cite{ibaz3}, in the context of $Z_3$ IIB orientifold compactifications. }  
pairs of
$H_u$, $H_d$ higgsinos $\&$ gauginos $\&$ three vector pairs of exotic SU(3) triplets in addition to two N=1
adjoint \footnote{Easily reproduced by the wrappings for the solution $(Z_a, Y_a) = (Z_b, Y_b) = (1, 1)$ at table (\ref{tabsu1}). } hypermultiplets [see eqn. (\ref{adj1})]; one particle in the adjoint of SU(3) (an octet) and one in the adjoint of SU(2)(a triplet).
We are using the central values \cite{naka}   
\beqa
M_z = 91.1876 \ GeV, & \alpha_{em}(M_z)^{-1} = 127.916 \nonumber\\
\alpha_3 (Mz)  = 0.1184 , & sin^2 ( \theta_Z ) = 0.23113 \ .
\label{dat1}
\eeqa
Depending on the scale at which, matter becomes massive and decouples from the Wilsonian 
effective action, we can envisage different unification scenarios. In all cases,  
the gauginos receive \cite{louis2} a typical mass of the order of the string scale \footnote{See also \cite{anastaso} for a recent discussion.} and there will be no contribution to (\ref{betas}) in all scenarios. 
\newline
\newline
$\bullet$ $\bullet$ {\bf 1st scenario : Non-susy string models with only the SM $\&$  three pairs of Higgses at low energies} 

\underline{{\bf 1st variant}  \ $n_H=6$}

\begin{figure}
    \centering
    \includegraphics[width=0.8\textwidth]{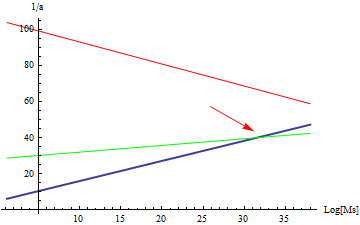}
    \caption{1st scenario : One loop gauge running of the couplings a$^{-1}$ as a function of $Log_{10}(E/GeV)$ for the non-susy MSSM model of table (\ref{give1}) at  $M_s \approx 5.03 \times 10^{13}$ \ GeV.
  The head of the arrow indicates the unification point.
The spectrum includes the SM $\&$ 3 pairs of Higgses at low energy. The unification point does not change if the higgses become one or three.}
\label{imageS} 
\end{figure}

In this scenario the higgsinos $\&$ vector triplets receive string scale masses 
from Yukawa\rq{}s in eq. (\ref{yuk}).  Let us also suppose, for the purposes of this section, that the adjoint matter receives string scale masses by some yet unknown mechanism.
 Then below $M_s$ and substituting $n_G =3$, $n_H = 6$ in eqn\rq{}s (\ref{betas}) 
\beq 
(b_1 , b_2 , b_3 ) = (\frac{23}{3}, -\frac{7}{3}, -7), \  B = \ \frac{44}{3}, \ \ a_2 (M_s) \ = \ a_3(M_s) \approx \  0.025 \ .
\eeq
The values of $a_2$, $a_3$ agree \footnote{We find $a_2 = 0.0252274..$, $a_3 = 0.0259045..$.} up to an error of 2.68 $\%$ as a result of using the experimental values of (\ref{dat1}) in the RG  eqn\rq{}s (\ref{RG}). 
Plugging them in (\ref{strings}), we find that the gauge couplings unify at the scale
\beq
 M_s \ = \ 5.03385 \cdot 10^{13} \ GeV   \approx 5.03 \cdot 10^{13} \ GeV ; B=\frac{44}{3}
\label{Aunik}
\eeq
as can be seen in figure (\ref{imageS}).

\underline{{\bf 2nd variant}  \ $n_H=1 \ or \ 3 $}

Suppose now that only one  
Higgs survives at low energy (as in the SM). In this case we get the usual SM contributions
\beq
n_H = 1, \ (b_1 , b_2 , b_3 ) = (\frac{41}{6}, -\frac{19}{6}, -7), 
\eeq
and $B = \frac{44}{3}$ as before and $a_2$, $a_3$  agree up to an error 12.2 $\%$; 
$a_2 = a_3 \approx 0.023$.  As the unification scale depends on experimental data (\ref{dat1}) and B,  
the value of the unification scale is the same as in (\ref{Aunik}). 
Notice that it is unlikely that only one Higgs could survive massless to low energy in the present models 
as the Higgses appear at a single intersection with multiplicity three (3). However, this is not the case within the SM-like non-sypersymmetric  models of \cite{kokos5}, \cite{kokos6}, \cite{louis2} based on toroidal IIA orientifolds, where the number of Higgses at low energy can be one.  
If the minimal number of Higgses at an intersection which survives to low energies, as in split supersymmetry models of table (\ref{give1}) this translates \footnote{} to 
\beq
n_H=3,   \ (b_1 , b_2 , b_3 ) = (\frac{43}{6}, -\frac{17}{6}, -7), \ B=\frac{44}{3}
\eeq
and $a_2 = a_3 = 0.023$ agree up to 8.3 $\%$. Thus with 
\beq
 SM \ \& \ \# higgs =1, 3, 6 \   \Longrightarrow \  M_s \ = \ 5.03 \times 10^{13} \ GeV  \ .
\eeq
Thus \lq\lq{}{\em in non-supersymmetric models coming from intersecting branes,  if the
SM survives at low energy below the string scale and the
 number of higgses is one, three or six, then always the unification scale of the strong and weak coupling occurs at the same point $M_s = 5.03 \times 10^{13}$ GeV (since the difference $-b_2 + b_1$ is in all cases equal to the constant  $10$ and $B=44/3$).} 
\newline
$\bullet$ $\bullet$ {\bf 2nd scenario : Non-susy string models with only the SM $\&$  three pairs of Higgses $\&$ minimal adjoint matter at low energies}

\begin{figure}
    \centering
    \includegraphics[width=0.9\textwidth]{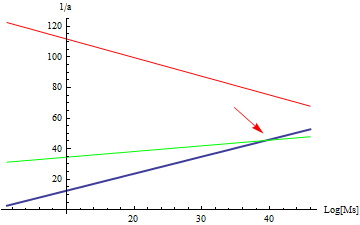}
    \caption{2nd scenario : Unification in the non-susy model of table (\ref{give1}). The spectrum includes SM $\&$ 3 pairs of 
Higgses $\&$ adjoint fermions (one SU(3) adj + one SU(2) adj). Ms $\sim 1.43 \cdot 10^{17}$ GeV.}.
\label{image2} 
\end{figure}

 In this case, we imagine a scenario where the extra triplets $C_1$, $C_2$ receive a mass of the order of the string scale from the vev
of $\nu_R$\rq{}s and the Higgsinos receive a Dirac mass term (assuming that it is of order $M_s$) as in (\ref{yuk}).
As a consequence, at low energies only the SM and three pairs of Higgses survive. The models also accommodate the non-chiral adjoint matter which preserves N=1 supersymmetry; one adjoint fermion from SU(3) and another one in the adjoint of SU(2) (\ref{nonma}). 
The $\beta$-functions become
$(b_1 , b_2 , b_3 ) = ( \frac{20}{3}+2,  -\frac{7}{3}+2, -7+2)= (\frac{23}{3} , -\frac{1}{3}, -5)$. 
 Then using (\ref{dat1}) in (\ref{strings}) 
\beq
a_3 = a_2  \approx  0.022,  \  \ \ B=\frac{34}{3}
\eeq
The values of $a_2$, $a_3$ agree up to an error of 4.35 $\%$ as a result of using the experimental values of (\ref{dat1}) in the RG  eqn\rq{}s (\ref{RG}). We find
\beq 
M_s = 1.43027 \cdot 10^{17} \  GeV \approx 1.43 \times 10^{17} \  GeV \ .
\label{unis} 
\eeq
The gauge  couplings $a_3$, $a_2$  unify at the scale $M_s$  as seen in figure (\ref{image2}).
\newline
$\bullet$ $\bullet$ {\bf 3rd scenario : Non-susy string models with only the SM $\&$ Higgses $\&$ adjoint matter $\&$ 3 pairs of Higgsinos at low energies}

If Higgsinos receive the suppressed mass (\ref{maxhigg1}) and survive to the electroweak scale then in addition to the matter of the 1st scenario, we may add the three pair of Higgsinos contribution. 
Then $(b_1 , b_2 , b_3 ) = ( \frac{29}{3} , -\frac{1}{3} +2 , -7+2 )$; +2 the contribution from the N=1 adjoint matter. As $B=\frac{34}{3}$ the unification scale does not change and is equal to (\ref{unis}).
\newline
$\bullet$ $\bullet$ {\bf 4th scenario : Non-susy string models with only the  SM $\&$ Higgsinos $\&$ 3 pairs of Higgses at low energy} \newline
Suppose that in the 3rd scenario, the adjoint matter is getting massive by an unknown mechanism. In this case we can neglect  the contribution +2 of the adjoint matter, thus  getting $(b_1 , b_2 , b_3 ) = ( \frac{29}{3}, -\frac{1}{3}, -7)$; $B=44/3. $ 
Unification occurs at the scale given in (\ref{Aunik}) as the value of B is the same in the 
1st scenario.

The 2nd $\&$ 3rd scenarios appear to be the most natural from the physical point of view, since
the high scale of unification, keeps the proton stable.

\section{\small Conclusions and outlook}

In this work, we have presented the systematic construction of the first four dimensional three family non-supersymmetric MSSM-like models that contain models which possess the spectrum of the MSSM fermions. The models explore the  
details of the construction presented in \cite{koz3z3}, where mostly GUT theories were described and where model constructions are described in terms of the \lq\lq{}effective\rq\rq{} wrappings $(Z, Y)$.  The Higgs scalars are coming from previously massive bosons that become tachyonic during electroweak symmetry breaking \cite{louis2, kokos5, kokos6}.
The intersection numbers that describe the number of particles at an intersection is always a multiple of three by construction.
The gauge group of the models is exactly $SU(3)_c \times
SU(2)_w \times U(1)_Y$. Any extra gauge group factors beyond the SM one, always are broken by available gauge singlets. 
 In its minimal construction, the particle content is made from the usual chiral SM matter and pairs of up, down higgsinos. Right handed neutrinos
 always appear, as always happens, at intersecting brane constructions. 
   Previous non-susy string models like the ones in toroidal orientifolds \cite{louis2, kokos5, kokos6}, have the baryon number a global gauged symmetry valid to low energies
and thus proton was stable (see also \cite{floko}).  
 In this respect, we examined gauge coupling unification in the simplest MSSM-like model made from 3-stacks of intersecting 
D6  branes, where the MSSM fermions live together with two N=1 SUSY adjoint multiplets, an octet of SU(3)$_c$ and a 3-plet of SU(2)$_W$. The most appealing scenario has a unification scale 
$\sim 10^{17} $ GeV, thus safeguarding the proton and avoiding dimension six operators \cite{axe}. It 
supports the existence of either the MSSM matter accompanied by adjoint matter up to $M_s$ or supermassive higgsinos ($\sim M_s$) and surviving to low energies SM matter and light adjoints.  Always, there are 3 pairs of higgses ${\tilde H}_u$, ${\tilde H}_d$ present.\newline
$\bullet$ Proton decay
\newline
The models are safe against proton decay. Take as an example the split susy models 
of table (\ref{give1}). The dimension six baryon number violating couplings \cite{wein} $Q_L Q_L Q_L L$ and 
$u_L^c u_L^c d_L^c e_L^{+}$ are not allowed in perturbation theory by charge conservation. They could be in principle generated by an M-instanton \cite{insta} with wrappings (X,Y) for which 
obeys 
\beq
I_{Ma}-I_{Ma^{\star}} = -1, \ I_{Mb}-I_{Mb^{\star}} = -1, I_{Mc}-I_{Mc^{\star}} = -1,  
\eeq  
For the a-, b- branes this requires that either $3(X-Y)=0, 3Y=1$  or $3(X-Y)=-1, X=Y$ conditions are satisfied. However there is no such soultion as X, Y take integer values.

We started our investigation for the construction of three generation N=0 string models 
by starting with the 
simplest construction that could accommodate the Standard Model gauge group that is
using three stacks. At 3- /5-stacks we found N=0 supersymmetric vacua 
with the chiral fermion
spectrum of the N=1 MSSM, with 3 species of right handed neutrinos and 3 pairs of chiral fermions
$H_u$, $H_d$ that play the role of the MSSM Higgsinos.  
These models break to only the SM at low
energy. 
In all constructions, symmetric/antisymmetric representations are present, and always the up-quark is massless at tree level; its mass may come through instanton corrections \cite{insta}, \cite{ibaura}, \cite{kokoinsta}.
\newline
$\bullet$ K-theory constraints 
\newline
The K-theory constraints, seen also as calculating the global gauge anomaly using a D-brane probe \cite{ura},
are related to the existence of the SU(2) global gauge anomaly when there is an odd number of D=4 
fermions charged in the fundamental representation \cite{wit}. 
 As in all of our models there is an even number of fermion doublets the models are 
K-theory anomaly free.  \newline
$\bullet$ Split Supersymmetry Models
\newline
Models considered in the last sections of our work, show 
us that in intersecting brane 
worlds it would be possible to construct models with the characteristics of split supersymmetry 
that not only can drive us to a successful prediction $sin^2 \theta =3/8 $ of the SU(5) GUT value for the 
Weinberg angle at the unification/sting scale $M_s$ but also incorporate light Higgsinos 
and string (GUT) scale gauginos. We found models [table (\ref{give1})] that the spectrum is minimal as it has only the SM and the higgsinos surviving massless below the string scale (with three pairs of Higgses and three right handed neutrinos) .  
This is promising against other constructions (even the existing N=1 one\rq{}s) where either the SM appears with incomplete spectrum or a lot of chiral massless matter remaining to low energies, as it can drives us closer to an exact implementation 
of split supersymmetry \cite{noti1} scenario in intersecting 
D-brane models by attempting to construct, in the future, models which will be the N=1 supersymmetric implementation of the present N=0 models, with the underlying N =1 SUSY, spontaneously broken to N=0.
In this case, important issues related the soft term structures and the constraint \cite{constraint} 
$tan \beta =1$, that appears to be consistent with the \lq\lq{}existence of one light Higgs\rq\rq{} condition may be examined, as non-supersymmetric string models from intersecting D6-branes obey it as well \cite{iba4}.
\newline 
$\bullet$ Dark matter, Gauge hierarchy, Unification
\newline
Nevertheless the models of this work they are possess dark matter candidates, since as suggested in \cite{dimo} \cite{mah} the neutral components of higgsinos in a small mixing with the binos in models where higgsinos are present [e.g those in table (\ref{give1})] can serve well this purpose. Furthermore  \cite{dimo},
if there are additional singlet fermions in the theory, with Yukawa
couplings to the Higgsino\rq{}s and the Higgs then the DM particle can then be an admixture of
the singlet and neutral Higgsino components. The latter case, could get realized in the 8-stack quasi-supersymmetric (QS) models of table (\ref{susymodels}) as the coupling $(H_u) (\langle {\tilde H}_d \rangle)^2 \langle S_1 \rangle )$ exists. In the latter models supersymmetry may be shown to be broken by a variation \cite{iba} 
of the angles between the branes. QS models with the split susy properties and
proton stability and a gauged baryon number appeared in \cite{floko}. 
In non-supersymmetric models one expects quadratic
loop corrections to the EW Higgs masses to appear as
the low energy manifestation of the gauge hierarchy problem. Indeed as QW models
possess N=1 supersymmetric sectors for particular choices of wrapping numbers, 
 quadratic Higgs divergences may cancel at one loop \cite{iba} (non-necessarily higher as it is not known). 
Thus a full solution to the gauge hierarchy at the weak scale remains
an open issue in the QS classes of split non-supersymmetric models. 
Finally, it is remarkable that in non-susy models from intersecting brane worlds, where 
$sin^2 (M_s) = 3/8$, the unification scale remains the same $5.03 \times 10^{13}$ when the number of surviving higgses is one (1), two (2) or three (3). 

\begin{center}
{\bf Acknowledgments}
\end{center}
I am grateful
to I. Antoniadis, G. Athanasiu, R. Blumenhagen, S. Dimopoulos, M. Douglas, N. Tracas,  
E. Floratos, G. Honecker, L. Ibanez, G. Kane, I. Klebanov, B. K\"ors, A. Lahanas, 
C. Munoz, T. Ott,  
and A. Uranga
for useful discussions. The author wishes to thank the High Energy groups
of Harvard, Princeton, Rutgers, Stanford, Kavli/UCSB, UCSB 
and Caltech for their warm hospitality where this work was completed.
This work was supported by 
the programme \lq\lq{}Pythagoras I\rq\rq{}.
\newpage
\begin{center}
{\bf Note added}
\end{center}
While the revised work of hep-th/0406258v3 was finishing in Nov. 2004 and were being prepared for 
submission we noticed \cite{noti1} that also proposed the existence  
of split supersymmetry 
scenarios in string theory. In fact the local SM brane quiver configuration
used in model A of \cite{noti1} is the one appearing within the global string spectrum configurations of sections (4.1) and (8.2) 
of this work.  The main bulk of this work, including the parts 1-7 appeared in hep-th in 
29th June 2004.  
 Since then, our
classification MSSM (SM+Higgsinos) stringy quiver structures appearing in tables  1, 3, 9,  have been also used 
by the authors of \cite{Cvetic2009} in their tables 4, 5, 7 respectively, in order to discuss instanton generation of missing 
mass couplings of matter in antisymmetric/symmetric 
representations in a local model context. Also our three stack MSSM quiver of table 1 and the four stack MSSM quiver of  table 9,   
have  been used in \cite{kiritsis} in their search for model quiver embeddings from Gepner models in sections  4.1, 4.2.6 of their work 
respectively. At the present hep-th/0406258v4, we have added new material in sections 7.1 and added the new sections 8, 10. 

\section{Appendix A: Wrapping $\#$\rq{}s for the minimal MSSM matter model}

Wrappings, subject to the interchanges (\ref{chawra1}), generating the SM's 
of table (\ref{tabold})  
 may be seen in 
tables (\ref{newa1}), (\ref{newa2}), (\ref{newa3}).
\begin{table}
[htb] \footnotesize
\renewcommand{\arraystretch}{1.3}
\begin{center}
\begin{tabular}{|c|c|}
\hline\hline
 $Brane$ & $\hspace{2cm}(n^1, m^1) \times (n^2, m^2) 
\times (n^3, m^3)$\hspace{2cm}\\
\hline\hline
$\{ a \}$   & $(0, \ 1) \times (1,\ 0)  \times (0,\ -1)$\\
\hline 
$\{ b \}$   & $(1,\ 0)  \times (1,\ 1) \times (-1, -1)$ \\
\hline
$\{ c \}$ & $(1,\ \ 1) \times (1,\  1)  \times (1,\ \ 0)$ \\
\hline
\hline
\end{tabular}
\end{center}
\caption{\small 
Wrapping numbers 
in the three stack D6-brane N=0 SMs of table (\ref{tabold}).
These wrappings come from the change $(n, m)_a \leftrightarrow  
(n, m)_b$ in the wrappings of 
table (\ref{newsu1}). 
\label{newa1} }
\end{table}

\begin{table}
[htb] \footnotesize
\renewcommand{\arraystretch}{1.3}
\begin{center}
\begin{tabular}{|c|c|}
\hline\hline
 $Brane$ & $\hspace{2cm}(n^1, m^1) \times (n^2, m^2) 
\times (n^3, m^3)$\hspace{2cm} \\
\hline\hline
$\{ a \}$   & $(0, \ -1) \times (0,\ 1)  \times (1,\ 0)$\\
\hline 
$\{ b \}$   & $(-1,\ -1)  \times (1,\ 0) \times (1, 1)$  \\
\hline
$\{ c \}$ & $(1,\ \ 0) \times (1,\  1)  \times (1,\ \ 1)$ \\
\hline
\hline
\end{tabular}
\end{center}
\caption{\small 
Wrapping numbers in the three stack N=0 SMs of table (\ref{tabold}).
These wrappings come from the change $(n, m)_a \leftrightarrow  (n, m)_c$ in 
the wrappings of table (\ref{newsu1}). 
\label{newa2} }
\end{table}

\begin{table}
[htb] \footnotesize
\renewcommand{\arraystretch}{1.3}
\begin{center}
\begin{tabular}{|c|c|}
\hline\hline
 $Brane$ & $\hspace{2cm}(n^1, m^1) \times (n^2, m^2) 
\times (n^3, m^3)$\hspace{2cm}\\
\hline\hline
$\{ a \}$   & $(1, \ 0) \times (0,\ -1)  \times (0,\ 1)$\\
\hline 
$\{ b \}$   & $(1,\ 1)  \times (-1,\ -1) \times (1, 0)$ \\
\hline
$\{ c \}$ & $(1,\ \ 1) \times (1,\  0)  \times (1,\ \ 1)$\\
\hline
\hline
\end{tabular}
\end{center}
\caption{\small 
Wrapping numbers in the three stack SMs of table (\ref{tabold}).
These wrappings come from the change $(n, m)_b \leftrightarrow  (n, m)_c$ in 
the wrappings of table (\ref{newsu1}). 
\label{newa3} }
\end{table}

\section{Appendix B : Wrapping $\#$\rq{}s for 5-stack MSSM-matter models}

In this appendix, we apply 
the exchanges (\ref{chawra1}) to the wrappings of table (\ref{susy1}).    
The resulting models have the same spectrum as the N=0 five stack Standard Models of 
table (\ref{newta0}). These choices of wrappings can be seen in tables
(\ref{susy2}), (\ref{susy3}), (\ref{susy4}). 

\begin{table}
[htb] \footnotesize
\renewcommand{\arraystretch}{1.4}
\begin{center}
\begin{tabular}{|c|c|}
\hline\hline
 $Brane$ & $\hspace{2cm}(n^1, m^1)(n^2, m^2)(n^3, m^3)$\hspace{2cm} \\
\hline\hline
$\{  a \}$   & $(1, \ 0)(1,\ 1)(-1,\ -1)$ \\
\hline
$\{ b \}$   & $(0,\ 1)(1,\ 0)(0, -1)$  \\
\hline
$\{ c \}$ & $(1,\ \ 0)(1,\ 1) (-1,\ \ -1)$  \\
\hline
$\{ d \}$ & $(1,\ 0)(1,\  1)( 1, \ 1 )$  \\
\hline
$\{ e \}$ & $(1,\ 1)(1,\  1)( 1, \ 0 )$  \\
\hline
\hline
\end{tabular}
\end{center}
\caption{\small
Wrapping numbers responsible for the N=0 five stack 4D three generation intersecting 
D6-brane SMs of table (\ref{newta0}).
These models are derived from table (\ref{susy1}) by the interchange $(n, m)_a \leftrightarrow (n, m)_b$.  
\label{susy2}}
\end{table}

\begin{table}
[htb] \footnotesize
\renewcommand{\arraystretch}{1.4}
\begin{center}
\begin{tabular}{|c|c|}
\hline\hline
 $Brane$ & $\hspace{2cm}(n^1, m^1)(n^2, m^2)(n^3, m^3)$\hspace{2cm} \\
\hline\hline
$\{  a \}$   & $(-1, \ -1)(1,\ 0)(1,\ 1)$  \\
\hline
$\{ b \}$   & $(0,\ -1)(0,\ 1)(1, 0)$   \\
\hline
$\{ c \}$ & $(-1,\ \ -1)(1,\ 0) (1,\ \ 1)$  \\
\hline
$\{ d \}$ & $(1,\ 1)(1,\  0)( 1, \ 1 )$ \\
\hline
$\{ e \}$ & $(1,\ 0)(1,\  1)( 1, \ 1 )$   \\
\hline
\hline
\end{tabular}
\end{center}
\caption{\small
Wrapping numbers responsible the N=0 five stack 4D three generation intersecting 
D6-brane SMs of table (\ref{newta0}).
These models are derived from table (\ref{susy1}) by the interchange $(n, m)_a \leftrightarrow (n, m)_c$.  
\label{susy3}}
\end{table}

\begin{table}
[htb] \footnotesize
\renewcommand{\arraystretch}{1.4}
\begin{center}
\begin{tabular}{|c|c|}
\hline\hline
 $Brane$ & $\hspace{2cm}(n^1, m^1)(n^2, m^2)(n^3, m^3)$\hspace{2cm}\\
\hline\hline
$\{  a \}$   & $(1, \ 1)(-1,\ -1)(1,\ 0)$ \\
\hline
$\{ b \}$   & $(1,\ 0)(0,\ -1)(0, 1)$    \\
\hline
$\{ c \}$ & $(1,\ \ 1)(-1,\ -1) (1,\ \ 0)$  \\
\hline
$\{ d \}$ & $(1,\ 1)(1,\  1)( 1, \ 0 )$ \\
\hline
$\{ e \}$ & $(1,\ 1)(1,\  0)( 1, \ 1 )$   \\
\hline
\hline
\end{tabular}
\end{center}
\caption{\small
Wrapping numbers responsible for the N=0 five stack 4D three generation intersecting 
D6-brane SMs of table (\ref{newta0}).
These models are derived from table (\ref{susy1}) by the interchange $(n, m)_b \leftrightarrow 
(n, m)_c$.  
\label{susy4} }
\end{table}

\newpage

\section{Appendix C: $L \leftrightarrow H_d$  5-stack MSSM matter model}

$\bullet$ {\bf Model C}

The spectrum of the new N=0 three generation models appearing in this appendix is 
derived from the
exchange (\ref{reta0}) on the five stack models of table (\ref{newta0}). 
The spectrum can be seen in
table (\ref{newmode0}). The Yukawa couplings for the quarks, leptons are 
given by
 \begin{table}
[htb] \footnotesize
\renewcommand{\arraystretch}{1.1}
\begin{center}
\begin{tabular}{|c|c|c|}
\hline\hline
\  \ \  \ ${\bf Matter\  \ for \  \ Y^1}$ & \ $Y^1$ 
\ & $(SU(3) \times SU(2))_{( Q_a, \ Q_b,\ Q_c,\ Q_d, \ Q_e )}\hspace{2cm}$  \\
\hline\hline
$\{ Q_L \}$  & $1/6$ & $3(3, {\bar 2})_{(-1, \ 1, \ 0, \ 0, \ 0)}$\hspace{2cm}  \\
\hline
$\{ u_L^c \}$ &  $-2/3$  & $3({\bar 3},1)_{(-2,\ 0,\ 0, \ 0, \ 0)}$\hspace{2cm}  \\
\hline
$\{ e_L^{+} \}$ & $ 1 $ & $3(1, 1)_{(0,\ 2,\ 0, \ 0,\  0)}$\hspace{2cm} \\
\hline
$\{ d_L^c \}$ & $1/3$ & $3(1, 1)_{(1, \ 0,\ 0,\ 1, \ 0 )}$ \hspace{2cm} \\
\hline
$\{ H_u \}$ & $1/2$ & $3(1, 2)_{(0, \ 1,\ -1, \ 0,\ 0 )}$ \hspace{2cm} \\
\hline
$\{ H_d \}$ & $-1/2$ & $3(1,{2})_{(0, \ -1,\ 0, \ 1, \ 0 )}$\hspace{2cm} \\
\hline
$\{ L \}$ & $-1/2$ & $3(1,{2})_{(0, \ -1,\ 0,\ 0, \ 1 )}$ \hspace{2cm} \\
\hline
$\{ S_3 \equiv \nu_L^c \}$ & $0$ & $3(1, 1)_{(0, \ 0,\ 1,\ 1, \ 0)}$\hspace{2cm} \\
\hline\hline
$\{ S_1 \}$ & $0$ & $3(3, 1)_{(0,\  0,\ 0,\ -2, \ 0  )}$\hspace{2cm} \\
\hline
$\{ S_2 \}$ & $0$ & $3(3, 1)_{(0,\  0,\ 0,\ 0, \ -2 )}$\hspace{2cm} \\
\hline
$\{ S_4  \}$ & $0 $  & $3(1, 1)_{(0, \ 0,\ 1,\ 0,\ 1 )}$\hspace{2cm} \\
\hline
$\{  S_5 \}$ & $ 0 $   & $6(1, 1)_{(0, \ 0,\ 0,\ -1, \ -1 )}$ \hspace{2cm} \\
\hline
\hline
$\{ X_1 \}$ & $-1/3$ & $3({\bar 3}, 1)_{(-1,\  0,\ -1, \ 0, \ 0)}$ \hspace{2cm} \\
\hline
$\{ X_2 \}$ & $1/3$ & $3(3, 1)_{(1,\  0,\ 0,\ 0, \ 1 )}$\hspace{2cm} \\
\hline
\end{tabular}
\end{center}
\caption{\small 
The three generation N=0 SM chiral spectrum from from five stacks of intersecting branes 
with its chiral spectrum with three pairs of Higgsinos.
On the top of the table the chiral structure of N=1 SM with right handed neutrinos.
 The middle 
part exhibits the extra gauge singlets while the bottom 
part includes the triplet exotics. These models can come from the models of 
table (\ref{newta0}) by the exchange (\ref{reta0}).
\label{newmode0} }
\end{table}
\beqa
Y_{(table \ \ref{newmode0})}\ =  \ 
 \lambda_{d} Q_L d_L^c H_d^H  S_5^H/M_s \ +  \ 
\lambda_{e} L e_L^{+} H_d^H S_5^H/M_s  
\ + \ \lambda_{\nu} L \nu_L^c H_u^H S_5^H /M_s  \ + \ &\nonumber\\
 X_1 X_2 ( \lambda_x^{(1)}  S_4^H S_2^H + \lambda_x^{(2)} S_3^H S_5^H )/M_s  \ , &\nonumber\\
\label{colwes1}
\eeqa
where the superscript $H$ denotes the tachyonic scalar superpartner of the
corresponding fermion.  
These models allow for a universal dependence of the masses of all the quarks and leptons on 
the tachyonic Higgs vev of the superpartner of $S_5$.
The fermions $H_u$, $H_d$ get massive by the following Yukawa terms 
\beq
 H_u H_d (\lambda_{\mu}^{(1)}  S_3^H S_1^H + \lambda_{\mu}^{(2)} S_4^H S_5^H )/M_s
\eeq

 \newpage

\section{Appendix D: $d^c_L \leftrightarrow X_2$  5-stack MSSM matter model}

$\bullet$ {\bf Model D}

The spectrum of the new N=0 three generation models appearing in this appendix is derived from the
exchange (\ref{exapen1}) on the five stack models of table (\ref{newta0}). It can be seen in
table (\ref{toraer1}). 
\begin{table}
[htb] \footnotesize
\renewcommand{\arraystretch}{1.1}
\begin{center}
\begin{tabular}{|c|c|c|}
\hline\hline
\  \ \  \ ${\bf Matter\  \ for \  \ Y^1}$ & \ $Y^1$ 
\ & $(SU(3) \times SU(2))_{( Q_a, \ Q_b,\ Q_c,\ Q_d, \ Q_e )}\hspace{2cm}$  \\
\hline\hline
$\{ Q_L \}$  & $1/6$ & $3(3, {\bar 2})_{(-1, \ 1, \ 0, \ 0, \ 0)}$\hspace{2cm}  \\
\hline
$\{ u_L^c \}$ &  $-2/3$  & $3({\bar 3},1)_{(-2,\ 0,\ 0, \ 0, \ 0)}$\hspace{2cm}  \\
\hline
$\{ e_L^{+} \}$ & $ 1 $ & $3(1, 1)_{(0,\ 2,\ 0, \ 0,\  0)}$\hspace{2cm} \\
\hline
$\{ d_L^c \}$ & $1/3$ & $3(1, 1)_{(1, \ 0,\ 0,\ 0, \ 1 )}$ \hspace{2cm} \\
\hline
$\{ H_u \}$ & $1/2$ & $3(1, 2)_{(0, \ 1,\ -1, \ 0,\ 0 )}$ \hspace{2cm} \\
\hline
$\{ H_d \}$ & $-1/2$ & $3(1,{2})_{(0, \ -1,\ 0, \ 0, \ 1 )}$\hspace{2cm} \\
\hline
$\{ L \}$ & $-1/2$ & $3(1,{2})_{(0, \ -1,\ 0,\ 1, \ 0 )}$ \hspace{2cm} \\
\hline
$\{ S_3 \equiv \nu_L^c \}$ & $0$ & $3(1, 1)_{(0, \ 0,\ 1,\ 1, \ 0)}$\hspace{2cm} \\
\hline\hline
$\{ S_1 \}$ & $0$ & $3(3, 1)_{(0,\  0,\ 0,\ -2, \ 0  )}$\hspace{2cm} \\
\hline
$\{ S_2 \}$ & $0$ & $3(3, 1)_{(0,\  0,\ 0,\ 0, \ -2 )}$\hspace{2cm} \\
\hline
$\{ S_4  \}$ & $0 $  & $3(1, 1)_{(0, \ 0,\ 1,\ 0,\ 1 )}$\hspace{2cm} \\
\hline
$\{  S_5 \}$ & $ 0 $   & $6(1, 1)_{(0, \ 0,\ 0,\ -1, \ -1 )}$ \hspace{2cm} \\
\hline
\hline
$\{ X_1 \}$ & $-1/3$ & $3({\bar 3}, 1)_{(-1,\  0,\ -1, \ 0, \ 0)}$ \hspace{2cm} \\
\hline
$\{ X_2 \}$ & $1/3$ & $3(3, 1)_{(1,\  0,\ 0,\ 1, \ 0 )}$\hspace{2cm} \\
\hline
\end{tabular}
\end{center}
\caption{\small 
The three generation N=1 SM from from five stacks of intersecting branes 
with its chiral spectrum with three pairs of Higgsinos and right neutrinos.
On the top of the table the chiral structure of N=1 SM. The middle 
part exhibits the gauge singlets while the bottom 
part includes the triplet exotics. These models can come from the models of 
table (\ref{newta0}) by the exchange (\ref{exapen1}).
\label{toraer1} }
\end{table}
Yukawa's for the quarks, leptons, and the $X_i$ triplets are given by
\beqa
Y_{(table \ \ref{toraer1})}\ =  \ 
 \lambda_{d} Q_L d_L^c H_d^H  S_2^H/M_s \ +  \
\lambda_{e} L e_L^{+} H_d^H S_5^H/M_s  
\ + \ \lambda_{\nu} L \nu_L^c H_u^H S_1^H/M_s  \ + \ &\nonumber\\
 X_1 X_2 ( \lambda_x^{(1)}  S_4^H S_5^H + \lambda_x^{(2)} S_3^H S_1^H )/M_s  
\ + \   H_u H_d (\lambda_{\mu}^{(1)}  S_4^H S_2^H + \lambda_{\mu}^{(2)} S_3^H S_5^H )/M_s &
\label{neasd11}
\eeqa
In fact, if  $\langle S_5^H 
\rangle   \approx   \langle S_2^H 
\rangle   \approx     \langle S_1^H 
\rangle  \approx M_s $, then the masses of the quarks and leptons 
depend universally on the scale of the electroweak symmetry breaking, assuming
$\langle H_u \rangle \approx \langle H_d \rangle $. 
The ``Higgsinos'' $H_u$, $H_d$, also get a non-zero mass from the last term in 
(\ref{neasd11}).

\end{document}